\documentclass[]{emulateapj}
\usepackage{natbib,amsmath}

\shorttitle{A Definitive Survey for Lyman Limit Systems}
\shortauthors{Prochaska, O'Meara, \& Worseck}

\begin{document}

\def\msun{M_\odot}
\def\hub{h_{72}^{-1}}
\def\umfp{{\hub \, \rm Mpc}}
\def\mzq{z_q}
\def\zabs{$z_{\rm abs}$}
\def\mzabs{z_{\rm abs}}
\def\msna{{\rm S/N}^{\rm A}_{912}}
\def\sna{S/N$^{\rm A}_{912}$}
\def\mnull{\nu_{\rm 912}}
\def\nnull{$\nu_{\rm 912}$}
\def\intl{\int\limits}
\def\nstatqso{469}
\def\numlls{192}
\def\maxoff{0.4}
\def\clls{1.9 \pm 0.2}
\def\alls{5.2 \pm 1.5}
\def\blls{-0.8 \pm 0.3}
\def\cmma{\;\;\; ,}
\def\perd{\;\;\; .}
\def\ltk{\left [ \,}
\def\ltp{\left ( \,}
\def\ltb{\left \{ \,}
\def\rtk{\, \right  ] }
\def\rtp{\, \right  ) }
\def\rtb{\, \right \} }
\def\sci#1{{\; \times \; 10^{#1}}}
\def \rAA {\rm \AA}
\def \zem {$z_{\rm em}$}
\def \mzem {z_{\rm em}}
\def \mzlls {z_{\rm LLS}}
\def \zlls {$z_{\rm LLS}$}
\def \mzend {z_{\rm end}}
\def \zend {$z_{\rm end}$}
\def \mzstrtt {z_{\rm start}^{\rm S/N^A=2}}
\def \zstrtt {$z_{\rm start}^{\rm S/N^A=2}$}
\def \zstrto {$z_{\rm start}^{\rm S/N^A=1}$}
\def \mzstrto {z_{\rm start}^{\rm S/N^A=1}}
\def \zstrt {$z_{\rm start}$}
\def \mzstrt {z_{\rm start}}
\def\smm{\sum\limits}
\def \lll  {$\lambda_{\rm 912}$}
\def \mlll  {\lambda_{\rm 912}}
\def \mtll  {\tau_{\rm 912}}
\def \tll  {$\tau_{\rm 912}$}
\def \mavtll  {\bar{\tau}_{\rm 912}}
\def \avtll  {$\bar{\tau}_{\rm 912}$}
\def \tigm  {$\tau_{\rm IGM}$}
\def \mtigm  {\tau_{\rm IGM}}
\def \kms  {km~s$^{-1}$}
\def \mkms  {{\rm km~s^{-1}}}
\def \lyaf {Ly$\alpha$ forest}
\def \Lya  {Ly$\alpha$}
\def \lya  {Ly$\alpha$}
\def \mlya  {Ly\alpha}
\def \Lyb  {Ly$\beta$}
\def \lyb  {Ly$\beta$}
\def \lyg  {Ly$\gamma$}
\def \ly5  {Ly-5}
\def \ly6  {Ly-6}
\def \ly7  {Ly-7}
\def \nhi  {$N_{\rm HI}$}
\def \mnhi  {N_{\rm HI}}
\def \lnhi {$\log N_{HI}$}
\def \mlnhi {\log N_{HI}}
\def \etal {\textit{et al.}}
\def \ob {$\Omega_b$}
\def \obh {$\Omega_bh^{-2}$}
\def \om {$\Omega_m$}
\def \ol {$\Omega_{\Lambda}$}
\def \gz {$g(z)$}
\def \mgz {g(z)}
\def \lyaf {Lyman--$\alpha$ forest}
\def \fnhi {$f(\mnhi,X)$}
\def \mfnhi {f(\mnhi,X)}
\def \myfnhi {f_{\rm Ly\alpha}(\mnhi,X)}
\def \yfnhi {$f_{\rm Ly\alpha}(\mnhi,X)$}
\def \lfnhi {$f_{\rm LLS}(\mnhi,X)$}
\def \dfnhi {$f_{\rm DLA}(\mnhi,X)$}
\def \sfnhi {$f_{\rm SLLS}(\mnhi,X)$}
\def \mlfnhi {f_{\rm LLS}(\mnhi,X)}
\def \mdfnhi {f_{\rm DLA}(\mnhi,X)}
\def \msfnhi {f_{\rm SLLS}(\mnhi,X)}
\def \ztot {$\Delta z_{\rm TOT}$}
\def \mztot {\Delta z_{\rm TOT}}
\def \mlplls {\ell_{\rm{PLLS}}(z)}
\def \lplls {$\ell_{\rm{PLLS}}(z)$}
\def \mlzlls {\ell_{\rm{LLS}}(z)}
\def \lzlls {$\ell_{\rm{LLS}}(z)$}
\def \lzo{$\ell_{\rm{\tau \ge 1}}(z)$}
\def \mlzo{\ell_{\rm{\tau \ge 1}}(z)}
\def \lzt{$\ell_{\rm{\tau \ge 2}}(z)$}
\def \mlzt{\ell_{\rm{\tau \ge 2}}(z)}
\def \ltlls {$\ell_{\tau \ge 2}(X)$}
\def \mltlls {\ell_{\tau \ge 2}(X)}
\def \llls {$\ell_{\rm{LLS}}(X)$}
\def \ldla {$\ell_{\rm{DLA}}(X)$}
\def \lzdla {$\ell_{\rm{DLA}}(z)$}
\def \lslls{$\ell_{\rm{SLLS}}(X)$}
\def \lzslls{$\ell_{\rm{SLLS}}(z)$}
\def \mlslls{\ell_{\rm{SLLS}}(X)}
\def \nlls {$n_{\rm LLS}$}
\def \slls {$\sigma_{\rm LLS}$}
\def \mnlls {n_{\rm LLS}}
\def \mslls {\sigma_{\rm LLS}}
\def \drlls {$\Delta r_{\rm LLS}$}
\def \mdrlls {\Delta r_{\rm LLS}}
\def \mlmfp {\lambda_{\rm mfp}^{912}}
\def \lmfp {$\lambda_{\rm mfp}^{912}$}
\def\kll{$\kappa_{\rm 912}$}
\def\mkll{\kappa_{\rm 912}}
\def \mbplls {\beta_{\rm pLLS}}
\def \bplls {$\beta_{\rm pLLS}$}
\def \btlls {$\beta_{\rm LLS}$}
\def \mbtlls {\beta_{\rm LLS}}
\def \ctlls {$C_{\rm LLS}$}
\def \mbtlls {\beta_{\rm LLS}}
\def \mlteff {\tau_{\rm eff}^\alpha}
\def \lteff {$\tau_{\rm eff}^\alpha$}
\def \teff {$\tau_{\rm eff}^{912}$}
\def \mteff {\tau_{\rm eff}^{912}}
\def \mnmin {\mnhi^{\rm min}}
\def \nmin {$\mnhi^{\rm min}$}
\def \O {${\mathcal O}(N,X)$}
\newcommand{\cm}[1]{\, {\rm cm^{#1}}}
\newcommand{\vfnhi}[1]{$f(\mnhi=10^{#1}\cm{-2},X)$}
\newcommand{\mvfnhi}[1]{f(\mnhi=10^{#1}\cm{-2},X)}
\def \snrlim {SNR$_{lim}$}
\def\mglls {{\gamma_{\rm LLS}}}
\def\glls {$\gamma_{\rm LLS}$}
\def\mavgt {<\mtll>}
\def\nplls {$N_{\rm pLLS}$}
\def\mnplls {N_{\rm pLLS}}
\def\bplls {$\beta_{\rm pLLS}$}
\def\mbplls {\beta_{\rm pLLS}}

\title{A Definitive Survey for Lyman Limit Systems
at $z\sim 3.5$ with the Sloan Digital Sky Survey}

\author{
J. Xavier Prochaska\altaffilmark{1}, 
John M. O'Meara\altaffilmark{2}, 
Gabor Worseck\altaffilmark{1} 
}
\altaffiltext{1}{Department of Astronomy and Astrophysics, UCO/Lick Observatory, University of California, 1156 High Street, Santa Cruz, CA 95064}
\altaffiltext{2}{Department of Chemistry and Physics, Saint Michael's College.
One Winooski Park, Colchester, VT 05439}

\begin{abstract}
We perform a semi-automated survey 
for $\mtll \ge 2$ Lyman Limit systems (LLSs) in
quasar spectra from the Sloan Digital Sky Survey, Data Release 7.
From a starting sample of 2473 quasars with $\mzem = 3.6 - 4.4$,  
we analyze \nstatqso\ spectra meeting strict seletion
criteria for a total redshift
path $\Delta z = 93.8$ and identify \numlls\ intervening 
systems at $\mzlls \ge 3.3$.
The incidence of $\mtll \ge 2$ LLSs per unit redshift, \lzt, is well
described by a single-power law at these redshifts:
$\mlzt = C_{\rm LLS} [(1+z)/(1+z_*)]^\mglls$,
with $z_* \equiv 3.7$, $C_{\rm LLS} = \clls$, and
$\mglls = \alls$ (68\%\ c.l.).
These values are systematically lower than previous estimates
(especially at $z<4$) but are consistent with recent measurements
of the mean free path to ionizing radiation.
Extrapolations of this power-law to $z=0$ are inconsistent with
previous estimations of $\ell(z)$ at $z<1$ and suggest a break
at $z \approx 2$, similar to that observed for the \lya\ forest.
Our results also indicate that the systems giving rise to LLS absorption
decrease by $\approx 50\%$ in comoving number density and/or physical size 
from $z=4$ to 3.3, perhaps due to an enhanced extragalactic ultraviolet background.
The observations place an integral constraint
on the \ion{H}{1} frequency distribution \fnhi\ and indicate that the
power-law slope $\beta \equiv d\ln\mfnhi/d\ln\mnhi$ is likely shallower
than $\beta = -1$ at $\mnhi \approx 10^{18} \cm{-2}$.  Including other
constraints on \fnhi\ from the literature, 
we infer that $\beta$ is steeper than $\beta = -1.7$ at 
$\mnhi \approx 10^{15}\cm{-2}$, implying at least two inflections in \fnhi.  
We also perform a survey for
proximate LLSs (PLLSs) and find that \lplls\ is systematically lower
$(\approx 25\%$) than intervening systems.  Finally, we estimate
that systematic effects impose an uncertainty of $10-20\%$ in the $\ell(z)$
measurements; these effects may limit the precision of all future surveys.

\end{abstract}


\section{Introduction}
Studies of hydrogen  absorption in the lines of sight towards 
distant quasars have served to both define, and in recent years bring
precision to, our cosmological models.  
The low density, highly ionized  \lyaf\ lines (a.k.a.\ the intergalactic
medium, IGM), with 
\ion{H}{1} column densities $\mnhi < 10^{17.2} \cm{-2}$,
have through their aggregate statistical properties (e.g.\ their
flux power spectrum, mean flux, and column density distributions)
constrained cosmological parameters such as the primordial power spectrum
and the baryonic mass density and astrophysical parameters like
the amplitude of the ionizing background
\citep[e.g.][]{rau98,cwb+03,mcdonald05,tytler04,fpl+08}.
The high-density, predominantly neutral damped \lya\ systems (DLAs),
with $\mnhi \ge 10^{20.3} \cm{-2}$, trace the gas which forms stars, and
likely represent the progenitors of modern-day galaxies 
\citep[e.g.][]{wlf+95,wgp05,pw09}.

The majority of \lyaf\ lines and the DLAs have, through
analysis of their \lya\ lines, precisely measured \nhi\ values
that permit detailed study of their physical properties (e.g.\ metallicity).
For systems with intermediate \nhi\ values ($\approx 10^{18} \cm{-2}$),
however, \lya\ and most of the Lyman series lines lie on the
flat portion of the curve-of-growth making the \nhi\ value difficult
to constrain.
On the other hand, these systems are optically thick to ionizing
radiation and impose a readily identified signature in a quasar
spectrum at the Lyman limit.
These so-called Lyman limit systems (LLSs), currently the least-well
studied of \ion{H}{1} absorption systems at high redshift,
are the focus of this manuscript.

Historically, the LLSs were among the first class of quasar
absorption line (QAL) systems to be surveyed \citep{tytler82}.
This is because their spectral signature is obvious
in low-resolution, low S/N spectra.  The principal challenge
is that the Lyman limit occurs redward of the atmospheric
cutoff only for systems with redshifts $z>2.6$.
For lower redshifts, one requires spectrometers on space-borne
ultraviolet satellites.
By the mid 1990's, samples of several tens of LLSs were generated
spanning redshifts $0 < z < 4$ \citep{ssb,lzt91,storrie94,key_lls}.
These results were derived from heterogeneous sets of quasars discovered
from a combination of color-selection, radio detection, and
slitless spectroscopic surveys.    The spectra, too, were acquired with a 
diverse set of instrumentation and therefore varying S/N and spectral
resolution mitigating differing sensitivity to the precise optical
depth at the Lyman limit.  Although the results were not fully
consistent with one another, the general picture that resulted
was a rapidly evolving population of absorption systems reasonably
described by a $(1+z)^{1.5}$ power-law.

Cosmologically, the LLSs contribute much if not most of the universe's
opacity to ionizing radiation.  And, until recently, the observed incidence
of the LLS provided the only direct means of estimating the mean
free path \lmfp\ at any redshift \citep[e.g.][]{mm93,mhr99,flh+08}.
In a companion paper \citep[][; hereafter PWO09]{pwo09}, we have presented a new
technique to measure \lmfp\ that circumvents any knowledge
of the LLSs.  A more precise census of the LLSs will serve as a 
consistency check for this \lmfp\ calculation, but is unlikely to
ever again be a competitive approach.  Instead, the incidence of LLS
can be used in combination with estimates of \lmfp\ to assess the
\nhi\ frequency distribution for gas 
with $\mnhi \approx 10^{16-18} \cm{-2}$, 
a regime that is very difficult to explore by studying individual
absorption systems.   Surveys of the LLSs are also likely to 
place tight constraints on $z \sim 3$
cosmological simulations that include radiative transfer.

Physically, the nature of systems that give rise to a LLS remains
an open question.  The systems with the largest \nhi\ values
(i.e.\ $\mnhi \ge 10^{19} \cm{-2}$, the so-called the super-LLS or SLLS and
DLAs) are likely associated with the interstellar medium and
outer regions of high $z$ galaxies.  These high \nhi\ systems, however,
are only a subset of the LLS population.  Unfortunately, a proper
modeling of the LLSs almost certainly requires careful modeling of 
radiative transfer in cosmological simulations
which has thus far been beyond the scope of
modern computations in cosmological simulations.  
Indeed, the few studies to date have tended
to severely underestimate the incidence of LLS 
\citep[][but see Kohler \& Gnedin 2007]{kwh+96,gkh+01}.
In recent simulations of high $z$ galaxy formation, however, 
theorists have placed great attention on `streams' of cold gas that
carry fresh material from the IGM to star-forming galaxies 
\citep{kkw+05,dbe+08}.  These cold streams have relatively large
hydrogen surface densities ($N_{\rm H} \sim 10^{20} \cm{-2}$) and could therefore
produce Lyman limit absorption provided the material has a
non-negligible neutral fraction.  Consequently, an accurate census
of the LLSs with redshift may directly constrain the nature and
prevalence of cold streams in the young universe.

A final, yet perhaps most important, motivation for studying the LLSs
is that these systems may dominate the census of metals at all epochs.
The majority of LLSs are metal-bearing, showing metal-line transitions of 
common low and high-ions \citep[e.g.][]{p99,pks2000}.
Because the estimated ionization corrections for LLSs with 
$\mnhi \approx 10^{18} \cm{-2}$ is large, observations of ions in an 
LLS likely track only a trace amount of the metals actually present 
in the gas.  Lyman limit systems may show a wider spread in their ionization and metal
content relative to the IGM or DLA, further emphasizing the need for a
robust LLS survey.

In this paper, we survey the homogeneous dataset of quasar
spectra from the Sloan Digital Sky Survey (SDSS), using all 
7 public data releases.  Our observational analysis
aims to produce the most precise measurement of 
the LLS incidence paying careful attention to systematic
biases.  The wavelength coverage and data quality of the SDSS
quasar spectra focus the survey at $z \approx 3.5$.  Future
work will depend on follow-up observations of well-defined
quasar samples at other wavelengths.

The paper is organized as follows.  In $\S$~\ref{sec:def},
we present a set of LLS definitions used throughout the manuscript.
The selection criteria and data quality of quasars from the SDSS database 
are described in $\S$~\ref{sec:quasars}.
The procedure to model the absorbed quasar continuum is presented in
$\S$~\ref{sec:continuum} and the search and characterization of LLSs is 
detailed in $\S$~\ref{sec:lls}.
The criteria used to measure the survey path are described
in $\S$~\ref{sec:path} and an assessment of systematic error and
bias from analysis of mock spectra is provided in $\S$~\ref{sec:mock}.
$\S$~\ref{sec:results} describes the principal results and the
implications for the IGM and cosmology are discussed in $\S$~\ref{sec:discuss}.
Finally, $\S$~\ref{sec:summary} presents a summary of the main findings.
Throughout the paper, we adopt a $\Lambda$CDM cosmology with 
$H_0 = 72 \, h_{72} \, \mkms \, \rm Mpc^{-1}$, 
$\Omega_{\rm m} = 0.3$, and $\Omega_\Lambda = 0.7$
and report proper lengths unless otherwise indicated.


\section{Lyman Limit System Definitions}
\label{sec:def}

The photon cross-section of a hydrogen atom at energies above the Lyman limit
may be approximated by:

\begin{equation}
\sigma_{\rm LL}(\nu \ge \mnull) \approx  
  6.35 \sci{-18} \ltp \frac{\nu}{\mnull} \rtp^{-3} \; \cm{2} \cmma
\end{equation}
with

\begin{equation}
\mnull = E_{912}/h = c/\mlll \cmma 
\end{equation}

\noindent and $E_{912} = 1$\,Ryd.  Specifically, 
$\mnull = 3.29 \sci{15}$\,Hz and 
$\mlll = 911.7641$\AA.
This implies an optical depth at wavelengths $\lambda \le \mlll$,

\begin{equation}
\tau_{\rm LL} (\lambda \le \mlll) \approx \frac{\mnhi}{10^{17.2} \cm{-2}} \ltp 
   \frac{\lambda}{\mlll} \rtp^{-3} \cmma
\end{equation}
where \nhi\ is the \ion{H}{1} column density. 
For a gas `cloud' intersecting a background source with intrinsic flux
$F_{\rm int}(\lambda)$, the observed flux $F_{\rm obs}(\lambda)$
blueward of the Lyman limit is 

\begin{equation}
F_{\rm obs}(\lambda \le \mlll) = 
  F_{\rm int}(\lambda) \exp \ltk -\tau_{\rm LL} (\lambda) \rtk
\label{eqn:flux}
\end{equation}

In what follows, we define a `standard' Lyman limit system to be
one where the optical depth at $\mlll$ is $\mtll \ge 2$, 
i.e.\ $\mnhi \ge 10^{17.5} \cm{-2}$.
We refer to these systems as the $\mtll \ge 2$ LLS.
This corresponds to greater than 85\%\ attenuation of an 
incident ionizing radiation field at $\nu = \nu_{912}$. 
By this definition, the class of LLS includes systems with
$10^{20.3} \cm{-2} \ge \mnhi \ge 10^{19} \cm{-2}$ 
\citep[the so-called super-LLS or sub-DLAs, 
hereafter referred to as SLLS; e.g.][]{opb+07} and systems with
$\mnhi \ge 10^{20.3} \cm{-2}$ \citep[the damped \lya\ systems, DLAs; e.g.][]{wgp05}.
In a few cases, we will distinguish between these `strong' LLSs from 
those with lower \nhi, referring to the latter as $\mtll \lesssim 10$ LLS.
We also note that our $\mtll \ge 2$ definition for a LLS differs
from other works which adopted $\mtll \ge 1$ or $\mtll \ge 1.5$.
These are all observationally-driven, not physically-motivated definitions.

Observationally, the absorption of a background source
by a $\mtll \ge 2$ LLS is 
readily apparent, even in low S/N spectra.
We define absorbers with $\mtll < 2$ (i.e.\ $\mnhi < 10^{17.5} \cm{-2}$) as the 
partial Lyman limit systems (pLLSs).  To survey these systems,
one requires higher quality spectra or an alternate approach to the analysis.

We define the redshift of an LLS as

\begin{equation}
\mzlls \equiv \frac{\mlll^{\rm LLS}}{\mlll} - 1
\end{equation}
where $\mlll^{\rm LLS}$ marks the observed onset of LL absorption.
In practice, this is is often estimated from strong Lyman series lines
(e.g.\ \lya, \lyb) that accompany the Lyman limit opacity.

We define the sub-set of LLSs that occur within 3000\kms\
of the emission redshift of the background source as proximate LLSs
(PLLSs).  We separate the analysis of these systems from the rest
to investigate changes in the incidence of optically thick gas
near high $z$ quasars due to, e.g.\ the quasar's radiation field
and local environment. 

Finally, we define the observable \lzt\ as the average
number of $\mtll \ge 2$ LLS detected per unit redshift at a given redshift.
In the previous literature, this quantity is also expressed
as $n(z)$, $dN/dz$, and $dn/dz$.  For comparison with previous results
in the literature, we also consider \lzo, the 
number of $\mtll \ge 1$ LLSs detected per unit redshift. 
We also attempt to separate the contributions to \lzt\ from
SLLSa \lzslls, DLAs \lzdla, and attribute the remainder
to the LLSs with $10^{17.5} \le \mnhi < 10^{19} \cm{-2}$, \lzlls.

\section{SDSS Quasar Sample and Spectroscopy}
\label{sec:quasars}

One of the primary objectives of the Sloan Digital Sky Survey (SDSS)
was to discover $\sim 100,000$ new quasars across the northern sky
\citep{yaa+00}.  The strategy of the SDSS team to achieve this
ambitious goal was a four-fold process:
(i) obtain deep, multi-band images across a large area of the sky;
(ii) select quasar candidates by demanding a point-like, point-spread-function
and imposing color criteria that separate the candidates from the
Galactic stellar locus;
(iii) obtain follow-up spectra for a magnitude-limited sample
with a fiber-fed spectrograph.
The details of target selection and quasar completeness with
redshift is described at length in a series of SDSS papers
\citep[e.g.][]{rfn+02}, but see \cite{wp09} for a new and more accurate analysis;  
and
(iv) automatically identify quasars and estimate their redshifts
(\zem) through template fitting to the optical spectroscopy.

Of these steps, the second has the greatest 
impact on a survey for high $z$ Lyman limit systems. 
The key issue for our survey is whether the presence 
of an intervening LLS biases the targeting of the
background quasar for follow-up spectroscopy.  
In effect, a high $z$ LLS severely `reddens' the quasar
at the bluest optical wavelengths of the SDSS imaging.  
With this effect in mind, the SDSS team imposed cuts
on the $(u-g)$ color which better separated the quasar
locus in color pace from the stellar locus.  The net effect,
however, is to bias the spectroscopic follow-up against 
quasar sightlines {\it without} a foreground LLS
(PWO09).  Our analysis indicates
an important bias for quasars with $\mzem < 3.6$. 
For this reason, we limit the statistical analysis
to quasars with $\mzem \ge 3.6$, but we also explore the bias
by considering the incidence of LLSs toward quasars with $\mzem = 3.4 - 3.6$.

The quasar spectra analyzed in this paper were taken from the Sloan
Digital Sky Survey, Data Release 7 \citep{sdssdr7}.  
We retrieved
the `best' 1D spectrum for every source flagged as a QSO or HIZ\_QSO.
This totaled 102,418 unique spectra\footnote{  
The SDSS spectra were processed through our automated algorithms for
finding absorption-line features and damped \lya\ candidates
\citep{phw05,shf06}.  
Approximately 10 of the spectra failed 
to be processed (primarily because the SDSS-reported emission redshift
is erroneous) and were removed from any subsequent analysis.}.
The SDSS survey employs a 
fiber-fed, dual-camera spectrometer that provides continuous wavelength coverage
from $\lambda \approx 3800-9200$\AA\ at a spectral resolution of 
FWHM~$\approx 150 \mkms$.
The SDSS team employs a custom, data-reduction pipeline that performs
sky subtraction using empirical measurements from fibers placed
to avoid objects detected in the SDSS images.
The majority of data suffer from excessive sky noise at long
wavelengths ($\lambda > 8000$\AA)
and the instrument throughput and atmospheric absorption
limits the sensitivity at the shortest wavelengths ($\lambda < 4200$\AA).

\begin{figure}
\centering
\includegraphics[width=3.5in]{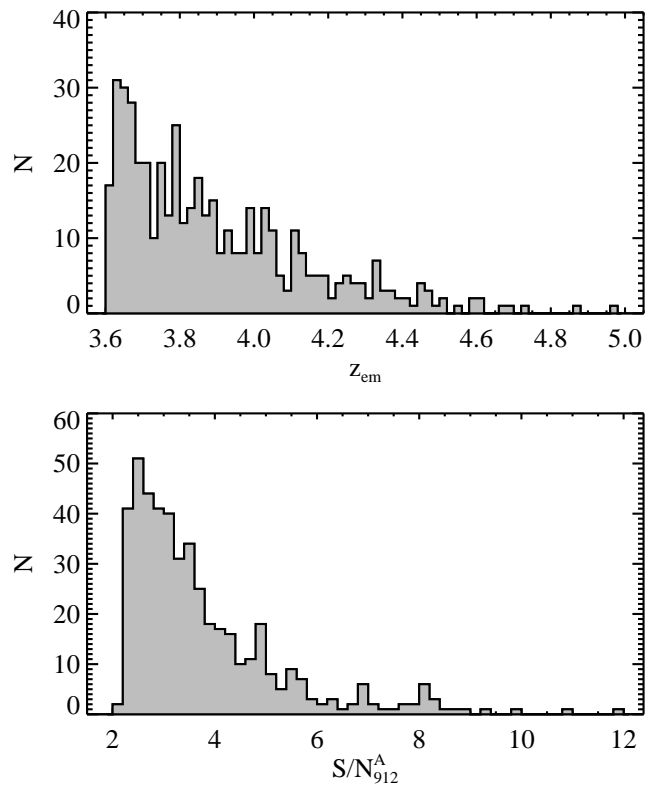}
\caption{Histograms of (top) the emission redshifts \zem\ for
the quasars comprising our survey and (bottom) the measurements
of \sna, the signal-to-noise of the absorbed continuum at the Lyman limit of
each quasar.  
The sample is restricted to \sna~$\ge 2$.
}
\label{fig:snz}
\end{figure}

A survey for \ion{H}{1} Lyman limit absorption in quasar spectra
involves two principle steps.  First, one must assess
the flux at wavelengths near the Lyman limit
in the quasar's rest-frame, $\lambda \lesssim \mlll(1+\mzem)$.  
We discuss our procedure for this step in the following section.  
The second step is to 
estimate the flux at wavelengths blueward of LLS candidates.
There are several characteristics of the SDSS spectroscopy which
negatively affect this estimate.  A generic concern is the poorer instrument
response at the bluest wavelengths.  At the bluest wavelengths,
many of the spectra exhibit a very low signal-to-noise ratio 
and yield flux estimates consistent with zero, even without an
intervening LLS.
Therefore, we have limited our survey to redshifts $\mzlls \ge 3.3$
corresponding to $\lambda > 3920$\AA.
In practice, we restrict the quasar sample to objects with
$\mzem \ge 3.6$ and perform a search for LLSs
at all redshifts, but then only analyze absorption systems with
$\mzlls \ge 3.3$.  We also limit the survey to quasars with
$\mzem \le 5$ because the SDSS spectra of higher redshift objects
are generally too low S/N to permit a robust analysis.
Figure~\ref{fig:snz}a presents histograms of the emission redshifts
for the statistical sample and the signal-to-noise of the
absorbed continuum at the Lyman limit
(\sna; see below for the definition).

Another difficulty with the SDSS spectra at blue wavelengths
is that the two-dimensional spectra of faint sources may be improperly traced.
On occasion, the 1D extractions
include flux from a neighboring object and yield a systematic 
overestimate of the flux.  This effect reduces the estimated opacity
for a LLS.  
Another issue, especially with a fiber-fed spectrometer, is that
the sky model is estimated from nearby fibers that are intentionally
placed on `object-free' regions of the sky.  Although the SDSS
project has worked carefully to mitigate the effects of variable
fiber efficiency,  significant misestimates of the sky are known to occur.  
We have identified tens of objects where the extracted
flux is significantly negative, indicating an overestimate
in the sky model.  
This may convert a partial LLS (with $\mtll < 1$) into
a $\mtll \ge 2$ LLS.  By a similar token, a proper $\mtll \ge 2$ LLS may
appear as a pLLS if the sky is underestimated.
We proceed under the expectation that this effect
is nearly random, i.e.\ for every underestimate of the sky there
is a corresponding overestimate, but this has not been rigorously
established.  
The SDSS fibers are sufficiently wide
(diameter of $3''$) that they will occasionally include flux
from a projected neighbor.  These coincident objects may be
much fainter than the quasar at redder wavelengths, but they
could contribute all of the flux blueward of a strong LLS
and lead to an underestimate of the LL opacity\footnote{An amusing
(and plausible) systematic effect related to this is contamination
by the light reflected from terrestrial satellites crossing the night sky.  
Even a brief `exposure' through the 3$''$ fiber could dominate the
flux at the bluest wavelengths, although this should generally be
mitigated by the fact that the SDSS team acquires 
3 unique exposures per target.}.

\section{The Absorbed Quasar Continuum}
\label{sec:continuum}

Absent any other sources of opacity, one can trivially
estimate \tll\ from
the quasar spectrum through measurements of the flux
both redward and blueward of the observed Lyman limit (Equation~\ref{eqn:flux}).  
In practice, however,
the quasar flux is also attenuated by line opacity from the so-called
\lya\ forest (a.k.a., the intergalactic medium; IGM).
For example, consider a Lyman limit system at $\mzlls = 3.5$
intervening a $\mzem = 4$ quasar.  The LLS attenuates the quasar flux
blueward of $\mlll^{\rm LLS} = 4103$\AA.  At this wavelength,
the quasar spectrum recorded on Earth will also include opacity
from the \lya\ forest at 
$z_{\rm \mlya} = (1+\mzlls)(\mlll/\lambda_{\rm Ly\alpha}) - 1$,
\lyb\ absorption from the IGM at 
$z_{\rm Ly\beta} = (1+\mzlls)(\mlll/\lambda_{\rm Ly\beta}) - 1$, etc.
It is necessary, therefore, to account for these additional sources
of opacity when estimating \tll.

We can express
the observed (rest-frame) quasar flux $F_{\rm obs}$ in terms of
the intrinsic flux (just) redward of the Lyman limit $F_{\rm int}$ as

\begin{equation}
F_{\rm obs} (\lambda \gtrsim \mlll) = F_{\rm int} \exp [-\mtigm(\lambda)] \cmma
\end{equation}
where \tigm\ is the effective opacity of the IGM from
Lyman series line-opacity\footnote{In the following, we do not
explicitly derive the 
opacity from metals in the IGM, but these may be considered included in \tigm.}.
The LLS introduces an additional, continuous opacity blueward of the
Lyman limit: 

\begin{equation}
F_{\rm obs} (\lambda \le \mlll) = 
  F_{\rm int} \exp [-\mtigm(\lambda) -\tau_{\rm LL}(\lambda)] \perd
\end{equation}
A precise estimate of $\mtll=\tau_{\rm LL}(\mlll)$, 
therefore, requires an estimation
of the absorbed quasar flux not its intrinsic flux.  Conveniently,
this quantity is the observed flux recorded in the spectrum at
$\lambda \gtrsim \mlll$.  
There are still significant challenges because
the IGM opacity is stochastic on both small (individual Lyman lines)
and modest scales (many 10\AA) 
and 
the intrinsic quasar spectrum (both shape and normalization) varies
from source to source.
We now describe an automated procedure
used to estimate the absorbed continuum from each quasar spectrum.

\begin{figure}
\centering
\includegraphics[height=3.5in,angle=90]{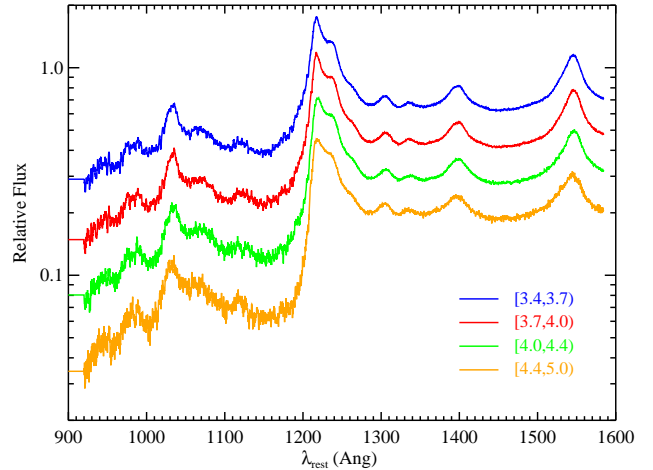}
\caption{Mean observed quasar spectra from 
the SDSS-DR3 in redshift intervals from
top to bottom of $z=[3.4,3.7); [3.7,4.0); [4.0,4.4)$ and
$[4.4, 5.0)$.    
The spectra have been offset by a factor of 1.5 for presentation
purposes and the normalization is not physically meaningful.
The increasing impact of the \lya\ forest is readily apparent
at wavelengths $\lambda_{\rm rest} < 1200$\AA.
At $\lambda_{\rm rest} < 920$\AA, we have set the template to a
constant value.  
}
\label{fig:template}
\end{figure}

The traditional method of estimating the
quasar continuum is to first identify
regions of unabsorbed quasar flux and then to interpolate a continuum level
between these regions.  For quasars at
high redshift, this method is particularly error prone, because at
wavelengths below \lya\ emission we expect few (and at very
high redshift, none) of the pixels to be free
of absorption from the IGM.   Moreover, this
traditional method frequently requires by-hand modification,
which is time-intensive and subjective to individual biases.
Methods do exist to automatically generate a quasar
continuum from emission-line characteristics \citep[e.g.][]{suzuki06},
but these are designed to infer the intrinsic quasar spectrum
not the IGM-absorbed continuum.

\begin{deluxetable*}{ccclccrrcc}
\tablewidth{0pc}
\tablecaption{SDSS-DR7 QUASAR SAMPLE\label{tab:qso}}
\tabletypesize{\footnotesize}
\tablehead{\colhead{Plate} &\colhead{MJD} & \colhead{FiberID} &\colhead{Object Name} &\colhead{\zem} & \colhead{$flg^a$} &
\colhead{A$^b$} & \colhead{B$^b$} & \colhead{Scale$^c$} & \colhead{\sna$^{,d}$}}
\startdata
 650&52143& 111&J000238.41$-10$1149.8& 3.938&0&$   2.60$&$  -0.29$& 1.11& 0.9\\
 750&52235& 608&J000300.34$+16$0027.7& 3.698&0&$  16.35$&$  -3.54$& 1.00& 1.5\\
 650&52143&  48&J000303.34$-10$5150.6& 3.646&0&$  46.77$&$ -11.08$& 0.96& 2.0\\
 750&52235&  36&J000335.21$+14$4743.6& 3.484&0&$  13.96$&$  -3.23$& 1.12& 0.8\\
 751&52251& 207&J000536.38$+13$5949.4& 3.686&0&$   9.96$&$  -2.00$& 1.00& 1.1\\
 751&52251& 562&J000730.82$+16$0732.5& 3.501&0&$   7.35$&$  -1.26$& 1.15& 1.2\\
 651&52141& 534&J001001.02$-09$0519.1& 3.720&2&\\
 751&52251&  39&J001115.23$+14$4601.8& 4.967&0&$  80.36$&$ -18.34$& 1.00& 3.0\\
 752&52251& 378&J001134.52$+15$5137.3& 4.325&0&$  15.67$&$  -3.60$& 1.00& 1.1\\
 752&52251& 204&J001328.21$+13$5828.0& 3.575&0&$  16.28$&$  -2.75$& 1.07& 2.5\\
 752&52251&   5&J001747.90$+14$1015.7& 3.955&0&$  21.69$&$  -5.06$& 0.92& 1.0\\
 753&52233& 310&J001813.88$+14$2455.6& 4.235&0&$  22.74$&$  -5.04$& 1.00& 1.7\\
 753&52233& 291&J001820.71$+14$1851.5& 3.936&0&$  70.03$&$ -16.86$& 1.32& 3.6\\
 753&52233& 391&J001918.43$+15$0611.3& 4.134&0&$   1.91$&$   0.13$& 1.13& 1.2\\
 390&51900& 271&J001950.05$-00$4040.7& 4.327&0&$  21.55$&$  -4.96$& 0.89& 1.9\\
\enddata
\tablecomments{{} List of all objects tagged as QSO or HIZ\_QSO in the SDSS Data Release 7 with estimated redshift $5 \ge z \ge 3.4$.}
\tablenotetext{a}{Flag indicating: 0=Normal; 1=Not at SDSS reported redshift and/or not a quasar (excluded); 2=Too low S/N for evaluation (excluded); 3=Strong BAL (excluded)}
\tablenotetext{b}{Absorbed continuum fitting parameters of the form: C = A + B $\log_{10} (\lambda/\rm \AA)$. We caution the reader
that these models are not meant to describe the intrinsic 
spectral energy distributions of the quasars (see the text).}
\tablenotetext{c}{Additinal scaling factor imposed by the authors on the best-fit absorbed quasar continuum.}
\tablenotetext{d}{Estimate of the signal-to-noise for the absorbed quasar continuum at $\lambda = \mlll (1+\mzem)$.}
\tablecomments{[The complete version of this table is in the electronic edition of the Journal.  The printed edition contains only a sample.]}
\end{deluxetable*}

Our approach is to match a template model of the average absorbed
continuum to each spectrum, allowing for a large-scale tilt 
(i.e. a unique underlying power-law slope) and arbitrary normalization.
We emphasize that this approach is not intended to recover the intrinsic
spectral energy distribution of the quasars.  Our scientific interest,
in this manuscript, is to model the absorbed
continuum of a quasar near its Lyman limit with the fewest number of
parameters.  Indeed, for some quasars we derive power-law slopes that
are likely unphysical due to the stochastic nature of IGM absorption, 
the non-linear effects of emission lines, intrinsic deviations from
a single power-law, and the narrow spectral range considered.  
We caution against using any of the following results
for studies on the physical properties of high $z$ quasars.

Our first step is to derive the templates for the average absorbed
quasar continuum.  Because the line-density of the IGM and therefore \tigm\
increase with redshift, we perform this analysis in small redshift intervals
($\delta z = 0.3$ to 0.6; Figure~\ref{fig:template}). 
For every quasar within the redshift interval,
we shift the spectrum to the quasar rest-frame 
using the SDSS reported \zem\ value.  
Next, we ``detilt" the spectrum by removing a power-law shape.
This power-law is determined as follows.  

We have
constructed from the SDSS-DR3 dataset an average template spectrum,
archived in XIDL\footnote{http://www.ucolick.org/$\sim$xavier/IDL}
as ``full\_SDSS\_LLS.fits''.
For each individual quasar spectrum,
we sample quasar pixels with wavelengths greater 
than \lya\ emission, divide by the template spectrum,
and measure the slope of the resulting spectrum.
After aligning the spectra in the quasar rest-frame
(nearest pixel), we median-combine the data in each redshift 
interval\footnote{This stack is not optimal for deriving \lya\
forest statistics \citep{dww08}.}.
The resultant template
represents the median intrinsic quasar continuum modulated by the median
flux decrement of the IGM.  Due to the presence of
Lyman limit systems, the quasar templates will not be useful at
wavelengths near and below the rest-frame Lyman limit;
this portion of the spectrum, however, can be used to 
constrain the mean free path to ionizing radiation (PWO09).
For wavelengths blueward of 920\AA, therefore,
we set the template to have the value recorded at 920\AA.
Conveniently, there are no strong emission features in the
quasar SED at these wavelengths \citep[e.g.][]{telfer02}.
The resultant template spectra are shown in Figure \ref{fig:template}.

With the templates constructed, a model of the absorbed continuum
for each quasar is determined as follows.
First, we shift the observed quasar spectrum to the rest-frame
and divide by the appropriate template spectrum (i.e.\ according to \zem).
Second, we sample the quasar in the wavelength range 
$950 \rm \AA < \lambda < 1800 \rm \AA$ and fit a power-law 
(p$(\lambda)$ = A + B log[$\lambda$/\AA]) to the observed 
flux, weighting by the inverse variance array.
The emission lines in this spectral range may bias the fit,
but we do review and modify these fits (see below).
The product of this power-law with the template, when
shifted to the observed frame, provides our model for the
absorbed quasar continuum.  The power-law parameters derived
in this fashion are listed in Table~\ref{tab:qso}.
Sample fits are shown in Figure~\ref{fig:ex_conti}.
With these models, we can calculate the ratio of the absorbed
continuum to the $1\sigma$-error array each quasar's Lyman limit,
which we denote as \sna.
Figure~\ref{fig:snz}b shows the distribution of \sna\ values
for the statistical survey.

\section{Lyman Limit System Search and Characterization}
\label{sec:lls}

In the following, we parameterize a LLS by two quantities:
(1) its absorption redshift \zlls\ and
(2) the total \ion{H}{1} column density \nhi.
Although the \ion{H}{1} Lyman series lines are sensitive to the component
structure and the Doppler parameters (also known as $b$-values)
of the `clouds' comprising an LLS,
the opacity blueward of \lll\ is insensitive to these details.
Furthermore, the SDSS spectra are generally of too poor quality
to constrain such structure using the observed Lyman series lines.
Therefore, our model of an LLS assumes a single cloud with a 
Doppler parameter of $b=30 \,\mkms$.
An implication of this parameterization is that two systems with
small redshift separation are modeled as a single system with
the total of the \nhi\ values.  Our tests with mock spectra ($\S$~\ref{sec:mock})
indicate that two absorbers with $\delta z < 0.1$ are often
indistinguishable from a single LLS.  The survey presented here,
therefore, refers to LL absorption smoothed over a redshift
interval of $\delta z \approx 0.1$.  We return to this point
in our presentation and discussion of the survey results.

We have developed an algorithm ({\it sdss\_findlls})
to automatically search for
and characterize LLS absorption in quasar spectra.
In brief, the code generates a set of model spectra
for the line and continuum opacity of
a single LLS with redshifts covering 
$z=z_0=(\lambda_0/\mlll-1)$ to (\zem+0.2) where $\lambda_0$
is the starting wavelength of the SDSS spectrum and \zem\ is the
quasar emission redshift reported in the DR7\footnote{Modified by
the analysis of \cite{phh08} where applicable for quasars from 
the DR5 release.}.
The grid of models assumes \nhi\ column densities 
$\log \mnhi = 16.0, 16.2, 16.4, ..., 19.8$
and a Doppler parameter $b = 30 \mkms$.  
We implement a grid with 0.2\,dex spacing in \nhi\ because very
few of the spectra have sufficient S/N to provide a more precise
estimate.  Furthermore, we estimate systematic uncertainties
(e.g.\ related to continuum placement, sky subtraction) to be of this order.
These models are convolved with the SDSS instrumental
resolution and then applied to the absorbed quasar continuum.
Finally, the code constructs a $\chi^2$ grid in \zlls\ and \nhi\
space, identifies the minimum $\chi^2$, and records the `best-fit' values.  
This approach differs from previous methods
which focused solely on the Lyman limit \citep[e.g.][]{storrie94} or
relied on `by-eye' analysis \citep{lzt91}.
In general, the spectra provide very little constraint
on \nhi\ for values exceeding $10^{17.5} \cm{-2}$ until the \lya\
profile becomes damped \citep[e.g.][]{phw05}.
Therefore, we report lower limits to \nhi\ for any LLS with 
$\mtll \ge 2$.  

For sightlines with a single LLS having $\mnhi > 10^{17.2} \cm{-2}$
and good \sna\ (i.e.\ greater than 5 pix$^{-1}$ 
at $\mlll^{\rm em} \equiv \mlll \times [1+\mzem]$),
we find that the automated
algorithm is highly successful on its own.
In practice, however, there are several aspects of the data and
analysis that require visual inspection of the spectra
and interactive modification to the model:
First, many spectra have 
such low S/N that \zlls\ and \nhi\ are poorly determined.
In these cases, a local minimum in $\chi^2$ can occur which
gives a mis-estimate of these quantities.
Second, the presence of multiple
absorbers along the sightline (e.g.\ one or more pLLSs with a lower
redshift LLS) gives a spectrum that cannot be well
modeled by a single LLS.
Third, a non-negligible number of the spectra
retrieved from the SDSS database purported to be high $z$
quasars are either at a lower redshift or are another class of
astronomical object altogether.  
Fourth, we found that half of the absorbed continuum models required
scaling to higher or lower value by greater than $10\%$.  
Finally,  we prefer to avoid quasars with 
with strong broad absorption line (BAL) or associated systems
to focus the analysis on intervening LLSs.

Given the above complications to an automated analysis, we
built a graphical user interface (GUI) within the IDL software
package ({\it sdss\_chklls}; bundled within XIDL)
that inputs the data and best-fit LLS model for each object.
Two of the authors (JXP and JMO) used this GUI to validate
and/or modify all of the models.  These authors 
flagged erroneous spectra (159 examples), 
strong BAL or associated absorption (quasars showing very strong
\ion{C}{4}, \ion{N}{5}, and \ion{O}{6} absorption; 290 quasars), 
or data with such low
S/N that any analysis was deemed impossible (114 spectra).
For the remainder of sightlines, the authors
could modify the continuum (via a multiplicative scalar; Table~\ref{tab:qso})
and/or change the model of LLS absorption (i.e.\ \zlls, \nhi).
This includes absorption due to candidate pLLSs.  
In many cases, \zlls\ was modified to correspond to the strongest,
local \lya\ absorption line at $\lambda = (1+\mzlls) \times 1215.67$\AA,
especially for those systems that also showed absorption at the
expected wavelength for \lyb.

After every sightline was analyzed in this manner,
the results from the two authors were compared to assess
consistency.  Roughly half of the spectra were reviewed
because of conflicts in the models.
The majority of these were associated 
with the absorbed continuum placement (typically offsets
of $5-10\%$) which implied 
differences in the search path of $|\Delta z| > 0.1$
(see $\S$~\ref{sec:path}).
In the majority of these cases, we simply averaged the
two estimations of the continuum.  
The second most frequent conflict was on the definition
of strong BAL absorption, primarily because we did not adopt
uniform or strict criteria.  In most cases, we 
conservatively excluded the sightline.
There were also $\approx 100$ cases where one author estimated
$\log \mnhi = 17.4$ when the other estimated   
$\log \mnhi = 17.6$, i.e., straddling the $\mtll=2$ boundary
that defines our LLS search. 
These were especially scrutinized for the presence of
higher-order Lyman series lines. 
Where necessary, the final best-estimate for \nhi\ was 
deferred to the third author (GW).

\begin{deluxetable}{lccccc}
\tablewidth{0pc}
\tablecaption{SDSS-DR7 INTERVENING $\mtll \ge 2$ LLS SURVEY\label{tab:survey}}
\tabletypesize{\footnotesize}
\tablehead{\colhead{Quasar} &\colhead{\zem} & 
\colhead{$z_{\rm start}^{\rm S/N=2}$} &\colhead{$z_{\rm start}^{\rm S/N=3}$} &
\colhead{\zlls}}
\startdata
\cutinhead{$\mzem < 3.6$}
J001328.21$+13$5828.0&3.575&  3.443 &\ldots&\ldots \\
J015741.56$-01$0629.6&3.564&  3.387 &  3.387 &3.387\\
J073947.17$+44$5236.7&3.575&  3.300 &\ldots&\ldots \\
J074914.13$+30$5605.8&3.436&  3.300 &  3.300 &\ldots \\
J075303.34$+42$3130.7&3.590&  3.300 &  3.300 &\ldots \\
J075859.81$+16$5811.8&3.439&  3.366 &\ldots&3.366\\
J080025.10$+44$1723.1&3.554&  3.300 &  3.362 &\ldots \\
J080525.53$+12$3438.7&3.425&  3.300 &  3.300 &\ldots \\
\cutinhead{$\mzem \ge 3.6$}
J001115.23$+14$4601.8&4.967&  4.567 &\ldots&\ldots \\
J001820.71$+14$1851.5&3.936&  3.536 &  3.596 &\ldots \\
J004219.74$-10$2009.4&3.880&  3.633 &  3.633 &3.633\\
J010619.24$+00$4823.3&4.449&  4.049 &  4.049 &\ldots \\
J011351.96$-09$3551.0&3.668&  3.615 &  3.615 &3.615\\
J012403.77$+00$4432.7&3.834&  3.434 &  3.434 &\ldots \\
J015048.82$+00$4126.2&3.702&  3.302 &  3.302 &\ldots \\
J015339.61$-00$1104.8&4.194&  3.879 &\ldots&3.879\\
\enddata
\tablecomments{The quasars with $\mzem < 3.6$ are not included in the final analysis.
The starting redshifts correspond to the wavelength at which the absorbed continuum model, starting from \zem, no longer exceeds the smoothed noise array by the specified \sna\ threshold.  This value is limited to a maximum offset
from \zem\ of $\delta z = \maxoff$.}
\tablecomments{[The complete version of this table is in the electronic edition of the Journal.  The printed edition contains only a sample.]}
\end{deluxetable}
 
\begin{deluxetable}{lccccc}
\tablewidth{0pc}
\tablecaption{SDSS-DR7 NON-STATISTICAL LLS SYSTEMS AND CANDIDATES\label{tab:llscand}}
\tabletypesize{\footnotesize}
\tablehead{\colhead{Quasar} &
\colhead{\zem} & 
\colhead{\zlls} & \colhead{log \nhi$^a$} &
\colhead{S/N$^b$} }
\startdata
J000238.41$-10$1149.8&3.938&3.809&17.2& 0.7\\
J000300.34$+16$0027.7&3.698&3.570&17.2& 1.3\\
J000303.34$-10$5150.6&3.646&3.467&17.6& 1.5\\
J000335.21$+14$4743.6&3.484&3.498&16.6& 0.9\\
J000536.38$+13$5949.4&3.686&3.580&19.0& 1.1\\
J000730.82$+16$0732.5&3.501&3.511&19.8& 1.3\\
J001115.23$+14$4601.8&4.967&3.995&17.8& 2.6\\
J001134.52$+15$5137.3&4.325&4.348&19.8& 1.1\\
J001328.21$+13$5828.0&3.575&3.282&19.8& 1.7\\
J001747.90$+14$1015.7&3.955&3.925&17.4& 1.0\\
J001813.88$+14$2455.6&4.235&4.151&17.2& 1.6\\
J001820.71$+14$1851.5&3.936&3.456&17.2& 2.4\\
J001918.43$+15$0611.3&4.134&4.053&17.8& 1.2\\
J001950.05$-00$4040.7&4.327&4.047&18.4& 1.8\\
J002120.05$+15$5125.7&3.698&3.671&17.8& 2.0\\
J002614.69$+14$3105.2&3.973&3.895&19.8& 1.3\\
\enddata
\tablenotetext{a}{The \nhi\ values listed serve as a rough estimate.  Typical uncertainties for systems with $\mnhi < 10^{17.5} \cm{-2}$ are at least 0.2\,dex.}
\tablenotetext{b}{Estimate of the S/N of the absorbed continuum (ignoring the effects of pLLS) at $\lambda = \mlll(1+\mzlls)$.}
\tablecomments{[The complete version of this table is in the electronic edition of the Journal.  The printed edition contains only a sample.]}
\end{deluxetable}

\begin{figure}
\centering
\includegraphics[width=3.5in]{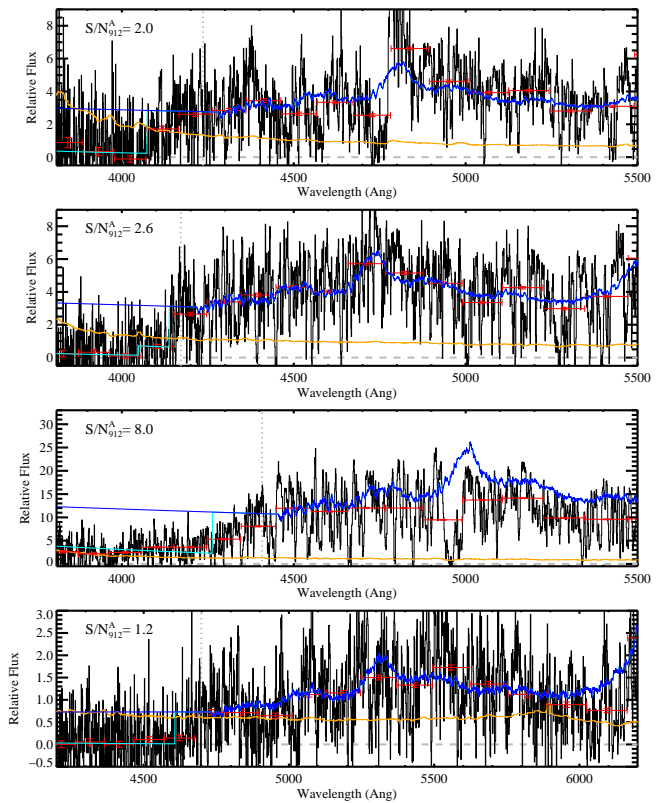}
\caption{Representative SDSS-DR7 spectra of quasars from our statistical survey.
Each example shows the absorbed continuum model (dark blue)
and the the model that includes LL absorption (cyan).  
The smoothed $1\sigma$ error array is given by the orange curve.
For this presentation, we have not shown the Lyman series in the model.
The vertical dotted line indicates the Lyman limit wavelength
in the quasar rest-frame, i.e.\ at $\lambda_{912}^{\rm em} = \mlll (1+z_{\rm em})$.
The red bins indicate the mean flux in intervals of 100 pixels.
The top panel shows a typical example of a $\mtll \ge 2$ LLS in data
that just satisfies our \sna\ criterion.
The next lower panel shows an example where we identify a 
pLLS candidate and then a $\mtll \ge 2$ LLS.  
The next lower panel shows a high \sna\ example whose quasar spectrum
rapidly declines with decreasing wavelength below $\lambda_{912}^{\rm em}$.
This absorbed continuum was modelled by including a pLLS candidate.
The lowest panel shows a low \sna\ example where a $\mtll \ge 2$ LLS
is apparent.
}
\label{fig:ex_conti}
\end{figure}

Tables~\ref{tab:survey} and \ref{tab:llscand} list the set of 
Lyman limit systems identified in the SDSS-DR7 for systems
that (respectively) influence our statistical analysis and otherwise. 
For each system, we list our best estimate for \zlls\ and the \nhi\ value.
Because the data offer minimal constraint on \nhi\
for values exceeding 10$^{17.5} \cm{-2}$,
in the statistical sample we only report a lower
limit to \nhi\ for all systems with best-estimates of
$\mnhi \ge 10^{17.5} \cm{-2}$. 
For the pLLSs, the typical $1\sigma$ uncertainty
is approximately 0.2\,dex for $\mnhi \ge 10^{17} \cm{-2}$
and is dominated by systematic error in
the continuum placement and the stochastic nature of IGM opacity. 
As noted above,
\zlls\ was frequently modified
in the interactive analysis to correspond to strong \lya\ and
\lyb\ lines.
Our analysis of mock spectra indicate typical redshift
uncertainties of $\sigma(z) < 0.02$ with occasional
`catastrophic failures' due to line-blending or spurious
spectra ($\S$~\ref{sec:mock}).
No attempt was made to improve this
estimate by searching for the presence of metal-line
absorption (e.g.\ \ion{C}{4}) outside the \lya\ forest.
There is a tendency, both in the automated algorithm and in 
interactive analysis, to set \zlls\ at the highest value possible
that can be accommodated by the data.  To this extent, we suspect
that there is a modest bias in our \zlls\ values to slightly
higher redshifts (less than 0.01 on average, but with
the occasional large offset).

Figure~\ref{fig:ex_conti} shows a representative sample of four quasar
spectra, zoomed into the region blueward of \lyb, with the
absorbed continuum and LLS models indicated. 
We provide snapshots of the 
LL region for all quasar spectra in the statistical sample 
online\footnote{http://www.ucolick.org/$\sim$xavier/SDSSLLS}.

\section{Survey Path}
\label{sec:path}

Analogous to galaxy surveys where one defines a search
volume based on the depth of imaging and spectroscopic
follow-up, measurements of the incidence of quasar absorption line systems
requires an estimate of the total spectral path sensitive 
to a robust search.  This is generally referred to as the redshift
path covered (by translating observed wavelength into redshift).
For the survey of LLSs, we have adopted the following criteria
for including spectral regions in the search.  These are based on our 
automated and interactive analysis of the SDSS spectra, 
our modeling of Keck/LRIS follow-up spectra, and our analysis
of simulated spectra ($\S$~\ref{sec:mock}, Appendix~\ref{appx:keck}):

\begin{enumerate}
\item The search path will begin at a minimum redshift of
  $z_{\rm start} \ge 3.3$
\item For the intervening LLS sample, the search path ends at the redshift
  $z_{\rm end}$
  corresponding to 3000~\kms\ (relativistic) blueward of the quasar
  redshift \zem.
\item The absorbed continuum flux must exceed twice the estimated error array, 
 i.e.\ \sna.
\item The search path begins at a maximum offset of 
   $\delta z = \mzem - z_{\rm start} \le 0.4$.
\end{enumerate}
The first criterion is motivated by the starting wavelength 
of the SDSS spectra ($\lambda_0 \approx 3800$\AA) and the
poorer quality of the data at the bluest wavelengths.
For redshifts less than 3.3, there is insufficient spectral
coverage and/or data quality to confidently assess the
presence of an LLS.  The second criterion minimizes
the influence of the quasar and its environment on
the analysis.  This criterion is relaxed in the study of PLLSs.

The third criterion is the most subjective, yet important,
for setting the redshift survey path of each quasar.
Algorithmically,  we impose this constraint by identifying
the first pixel blueward of $\mlll^{\rm em} \equiv \mlll(1+\mzem)$
where our model of the absorbed quasar continuum falls
below twice the median-smoothed (15 pixels) $1\sigma$ error array. 
This pixel defines the starting redshift \zstrtt\ corresponding
to a \sna=2 limit.  
If the first pixel blueward of $\mlll^{\rm em}$ does not satisfy
the S/N threshold, the quasar has zero redshift path, 
i.e.\ $z_{\rm start} = z_{\rm end}$.
One can, of course, define 
starting redshifts
corresponding to higher (or lower) \sna\ limits;  indeed, our
fiducial choice of \sna=2 should be considered arbitrary, although 
it is guided by our analysis of real and simulated spectra.
And to avoid a systematic bias associated with pLLSs (see $\S$~\ref{sec:pLLS}),
one must choose the \sna\ criterion to be sufficiently high
to discover $\mtll \ge 2$ LLS even in the presence of a 
pLLS\footnote{Contrary to some of the earliest work 
on LLSs \citep[e.g.][]{tytler82},
we do not use survival statistics to estimate the number of LLSs
at redshifts below any observed LLS.}.
We investigate the impact of this choice on our results
later in the manuscript. Finally, our fourth criterion is imposed to
mitigate the cumulative effects that pLLSs can have on our ability to
detect LLS with $\mtll \ge 2$.  That is, trials with mock spectra
($\S$~\ref{sec:mock}) indicate that mulptiple pLLSs along a single
sightline may prevent the detection of a $\mtll \ge 2$ LLS
and that this bias is minizmized provided $\delta z < 0.4$.
Furthermore, we find that the extrapolation of the absorbed
continuum from the quasar's Lyman limit often is a poor
model for $\delta z > 0.4$.

The starting redshift is further modified by the presence 
of Lyman limit absorption.  In the case of a $\mtll \ge 2$ absorber,
the quasar flux is severely depressed below $\lambda = \mlll (1+\mzlls)$
and we terminate the search path at this wavelength.  
Specifically, this implies $z_{\rm start} \ge \mzlls$ for all
sightlines with an $\mtll \ge 2$ LLS.
For sightlines where one or more pLLSs are identified\footnote{Again,
all pLLSs should be considered candidates.  Some of the spectral
features modeled as pLLSs are instead due to unusual variations
in the absorbed quasar continuum (e.g.\ highly reddened quasars).}, 
we had originally intended to terminate the search once the absorbed 
quasar continuum convolved with the pLLS absorption failed to satisfy
the S/N threshold.  In our analysis of mock spectra, 
however,  we found that this introduces a ``pLLS-bias'' where the
incidence of $\mtll \ge 2$ LLS is overestimated ($\S$~\ref{sec:pLLS}).
In part, our \sna=2 threshold is chosen so that one can 
robustly search for $\mtll \ge 2$ LLSs even along sightlines where
one or more pLLSs are present.  

Table~\ref{tab:survey} presents the list of quasars in SDSS-DR7
that 
(i) have $3.6 \le \mzem \le 5$, 
(ii) were not identified to
exhibit strong BAL signatures, and 
(iii) have $3.3 \le \mzstrtt < \mzem$.
There are \nstatqso\ quasars satisfying these criteria.
For each sightline, we list the starting redshifts for
\sna=2 and 3 limits.
We report results for these two values of the S/N threshold to
search for data-quality biases. 
Table~\ref{tab:survey} also lists all Lyman limit systems
with $\mzlls \ge \mzstrtt$.  
We do list these quantities for quasars with $\mzem < 3.6$,
but these were not included in our final statistical analysis
because of the bias related to SDSS targeting criteria
previously mentioned by PWO09.

Using the values presented in Table~\ref{tab:survey}, it is
straightforward to calculate \zend\ and the redshift search
path for each quasar: $\Delta z_i = \mzend - \mzstrt$.  For
a given \sna\ limit, the total search path for the full dataset is 

\begin{equation}
\Delta z_{\rm TOT} = \smm \Delta z_i  \perd
\end{equation}
We calculate $\mztot^{S/N=2} = 96$.
To maintain a homogeneous sample and set of search criteria,
we do not include previous studies in our analysis.
We compare against previous results in $\S$~\ref{sec:discuss}.

\begin{figure}
\centering
\includegraphics[height=3.5in,angle=90]{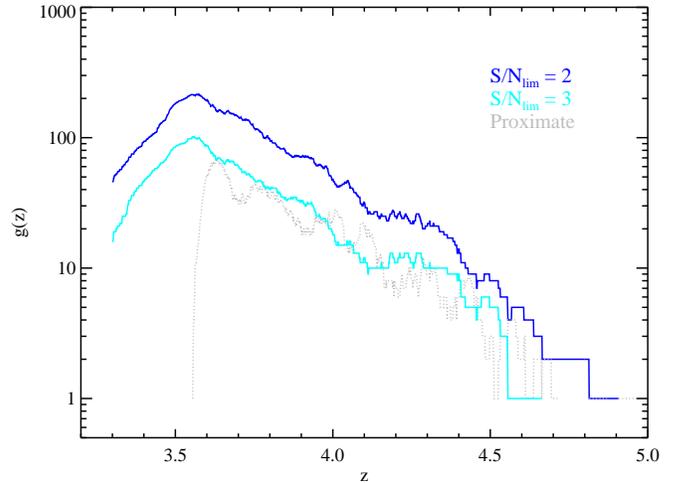}
\caption{Summary of the redshift path surveyed in the SDSS-DR7
for $\mtll \ge 2$ LLS absorption assuming \sna\ thresholds of 2 and 3
(solid curves).  Specifically, $g(z)$ represents the number of 
unique quasars in the SDSS that provide a search for LLSs over
the interval $dz=0.001$ at redshift $z$.  The dotted curve, meanwhile,
represents the same quantity but for proximate LLSs (PLLSs; systems
within $\delta v = 3000 \mkms$ of the quasar) and corresponds
to a \sna=2 threshold.  This curves fall rapidly for $z<3.6$
because we limit the analysis to quasars with $\mzem \ge 3.6$.
The non-monotonic nature of $g(z)$ is due primarily to the presence
of LLSs along the quasar sightlines which stochastically 
truncate the search path.
}
\label{fig:goz}
\end{figure}

In Figure~\ref{fig:goz}, we present the sensitivity function
\gz\ which expresses the number of SDSS quasars at redshift $z$
where a robust search for LLSs is possible.
The several solid curves represent differing \sna\ limits for the survey.
We also present the sensitivity function for proximate LLSs (PLLSs)
which corresponds to the sample of quasar spectra that satisfy
the $\msna \ge 2$ criterion at 3000~\kms\ blueward of \zem.

With the definition of the search path 
and the identification of the Lyman limit systems along each
sightline ($\S$~\ref{sec:lls}),
it is straightforward to calculate the incidence of intervening
$\mtll \ge 2$ LLS per redshift interval, \lzt.  
The standard estimator is to compare the total number of LLSs
against the total survey path in discrete redshift intervals.
We will return to evaluate \lzt\ and discuss the values
after exploring several sources of systematic error.

\section{Mock Spectra and Systematic Errors}
\label{sec:mock}

With a survey the size of SDSS, one can quickly reduce the statistical
noise in measurements to very small levels.  In this
regime, one must carefully assess all sources of systematic
error as these may dominate the measurement uncertainty.
To this end, we have conducted a range of tests with mock
spectra as described in this section.  The casual reader
may wish to skip to the summary of this section ($\S$~\ref{sec:mocksumm})

\subsection{Mock Spectra Construction}

We generated a set of 800 mock SDSS quasar 
spectra and analyzed them in the same way as the real data in 
order to assess bias and completeness in our LLS survey. 
The \ion{H}{1} forest absorption spectra 
were generated via a Monte Carlo routine similar 
to the one described in \cite{dww08}, 
assuming that the \lya\ forest is well characterized by 
three independent distributions: 
(i) the \lya\
line-incidence $\ell_\alpha (z) \propto (1+z)^\gamma$, 
(ii) the \ion{H}{1} column density distribution $f(\mnhi) \propto \mnhi^\beta$, 
and (iii) the Doppler parameter distribution parameterized as 
$f(b) \propto b^{-5} \exp[-b_\sigma^4/b^4]$ \citep{hr99}. 
Each simulated line of sight was filled with \ion{H}{1} \lya\ absorption lines at 
$2<z<4.6$ until the \ion{H}{1} \lya\ effective optical depth was consistent with 
\cite{fpl+08}, both in normalization and redshift evolution. We did not model the 
$z\sim 3.2$ dip in the effective optical depth 
measured by Faucher-Giguere et al., 
and instead adopted a simple power-law 
$\tau_{\rm eff}^{\alpha} =0.0011(1+z)^{4.23}$. 
If the number of lines in a given patch of the forest is 
Poisson-distributed, a power-law line density evolution 
$\ell_\alpha (z) \propto (1+z)^\gamma$ yields a power-law 
effective optical depth evolution $\tau_{\rm eff}^\alpha
\propto (1+z)^{\gamma+1}$ \citep{zuo93}. 
The column density distribution was modeled with a single power-law index 
$\beta=-1.5$ for $12<\log(\mnhi)<19$, but with a 0.5\,dex break at 
$\log(\mnhi)=14.5$ in order to account for the dearth of high column density lines, 
consistent with observations 
\citep[e.g.][and our own inferences, $\S$~\ref{sec:fnhi}]{hkc+95,kim02}. 
For the Doppler parameter distribution we set 
$b_\sigma=24 \, \mkms$ \citep{kim+01}. 

Because SLLSs $(19 \le \log(\mnhi)<20.3)$ and DLAs $(\log(\mnhi) \ge 20.3$) have 
different column density distributions and are usually excluded in 
measurements of $\tau_{\rm eff}^\alpha$,
these were added after the $\log(\mnhi)<19$ 
line forest converged to the chosen $\tau_{\rm eff}^\alpha (z)$. 
To constrain the redshift evolution of SLLSs, we combined the sample by 
\cite{opb+07} and the lower limit given in \cite{rtn06}, 
yielding $\ell_{\rm SLLS} (z) \sim 0.066(1+z)^{1.70}$. 
For the SLLS column density distribution we adopted $\beta=-1.4$ 
\citep{opb+07}. 
The DLAs were modeled via $\ell_{\rm DLA} (z) =0.044(1+z)^{1.27}$ \citep{rtn06} 
and $\beta=-2$ \citep{phw05}, 
ignoring deviations in \fnhi\ from a single power law. 
The Doppler parameter distribution was left unchanged.

With the overall opacity of the modeled \lya\ forest consistent with observations, 
and the high column density systems incorporated, we used the generated line lists to compute \ion{H}{1} Lyman series (up to Ly30) and Lyman continuum absorption spectra. In total we computed 800 different lines of sight for quasars in the redshift range of our sample, 
160 each at $z=3.4, z=3.6, z=3.8, z=4.0$ and $z=4.2$, respectively. 
From these we then generated mock SDSS spectra. First, the resolved \ion{H}{1} forest spectra were multiplied onto synthetic quasar SEDs generated from principal component spectra \citep{suzuki06}. 
We then degraded the resolution of the mock spectra to $R=2000$ by convolving them with a Gaussian, and rebinned them to $\delta v=69 \mkms$, matching the approximate 
resolution and pixel size of the SDSS spectra. Finally, we added Gaussian noise to the mock SDSS spectra. In each mock spectrum, 
the S/N was normalized in the quasar continuum 
at 1450\AA\ and varied as a function of flux and wavelength according 
to the throughput of the SDSS spectrograph. The sky level was 
approximated as a constant and readout noise was also incorporated. 
Finally, for a subset of the mocks
we imposed a sky subtraction error implemented by subtracting/adding
a constant to the spectrum.
These were generated to mimic such systematic errors
that occasionally occur in the SDSS spectra.

\begin{figure}
\centering
\includegraphics[height=3.5in,angle=-90]{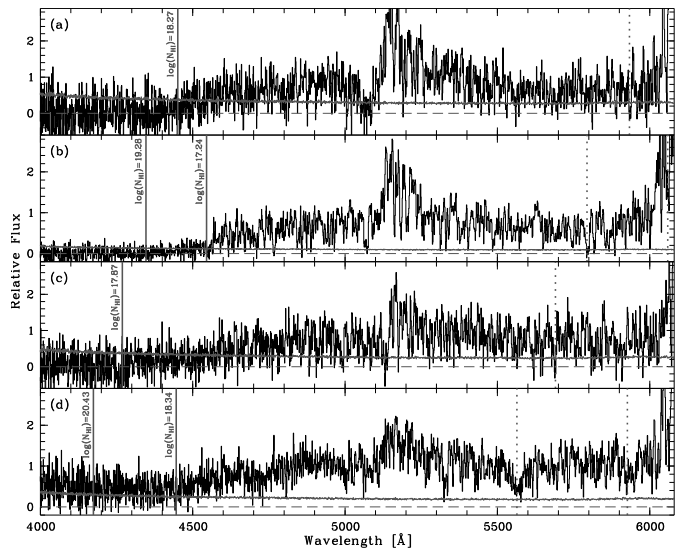}
\caption{
This figure presents a sample of the mock spectra analyzed in the paper.
(a): A spectrum representative of the typical data quality 
(\sna$\sim 2$ at the Lyman limit of the quasar). 
There is a LLS ($z=3.88059, \log\mnhi=18.27$) that can be easily identified 
at this \sna.
Two systems at higher $z$ with $\mnhi \approx 10^{16} \cm{-2}$
modify the continuum but do not produce a $\mtll \ge 2$ LLS;
(b): A high-S/N spectrum (\sna$\sim 6$ at the Lyman limit of the quasar), 
rendering its partial LLS ($z=3.98387, \log\mnhi=17.24$) 
easily visible. An additional SLLS at lower redshift 
$(z=3.76609, \log\mnhi=19.28)$ sets the flux to zero at the Lyman limit.
(c): 
This mock spectrum shows a slow `roll-off' in flux blueward of the quasar's
Lyman limit.
There is a LLS $(z=3.68084, \log\mnhi=17.87)$, 
and five systems with $16 < \log \mnhi <17$ at higher $z$ that produce 
the roll-off. The S/N at the Lyman limit of the quasar is quite low 
(S/N$\sim 1.5$) and the intervening systems further decrease 
the S/N, rendering the LLS invisible.
(d): A spectrum with strongly undersubtracted sky background. The 
first strong system encountered is a LLS 
$(z=3.87495, \log\mnhi=18.34)$, but the flux is above zero even 
after hitting a DLA $(z=3.57678, \log\mnhi=20.43)$. 
One can assess that the sky subtraction is poor, because 
the \lya\ profile of the DLA does not saturate. 
The solid/dotted veritcal lines in each panel trace the lyman limit/\lya\
line of each absorber.
}
\label{fig:mockex}
\end{figure}

Four representative examples are shown in Figure~\ref{fig:mockex}. 

\begin{figure}
\centering
\includegraphics[width=3.5in]{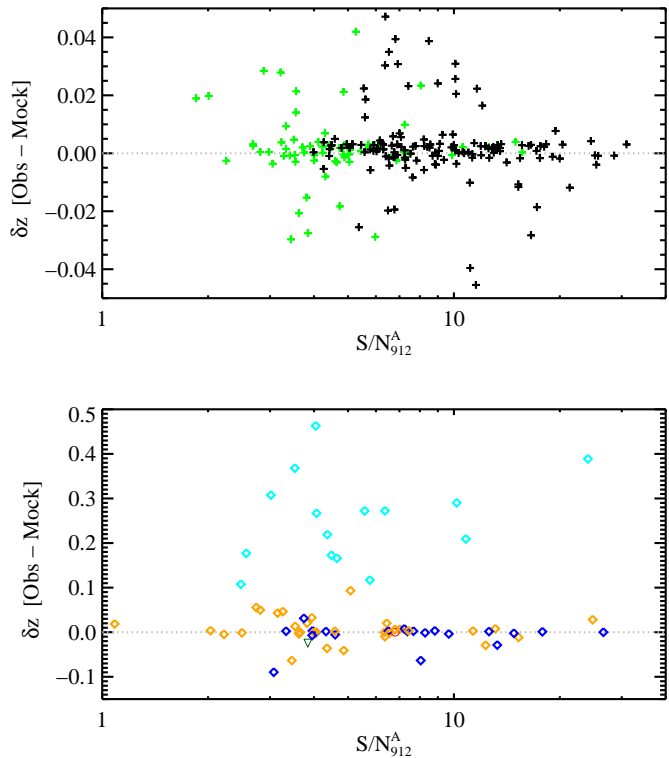}
\caption{Offset in redshift $\delta z$ for the `observed' LLS and pLLS
from the \zabs\ value of the nearest `true' $\mtll \ge 2$ LLS
in our mock spectra.  The analysis is restricted to the highest
\zabs\ LLS along each sightline.
In the upper panel, we show the $\delta z$ value for sightlines
where a LLS was `observed' and actually exists.  
Lighter/darker (green/black) points correspond to 
LLSs discovered outside/within the statistical survey path (i.e.\ \sna=2).
These are the majority of cases $(>80\%$)
and we find small $\delta z$ values with a small, but important
bias to $\delta z > 0.01$.
In the lower panel, we show false negatives (diamonds) and false
positive (triangles) detections.  The former correspond to a true LLS
that was observed as ($\delta z < 0.1$) or hidden by a pLLS ($\delta z > 0.1$).
The misidentifications ($\delta z < 0.1$) are dominated by the mock
spectra with large underestimates in the sky background.  The dominant
effect of a sky subtraction bias is an underestimate in the incidence
of LLSs.  For the misidentifications with $\delta z < 0.1$, the dark points
indicate systems that satisfy all of the survey criteria 
(e.g.\ $z > 3.3$, \sna~$\ge 2$).
}
\label{fig:mockdz}
\end{figure}

\subsection{LLS Recovery and Sky Subtraction Bias}

Two of the authors (JXP,JMO) analyzed mock spectra using
the identical tools and procedures applied to the SDSS spectra;
these steps were done without knowledge 
of the mock line distribution and column densities (constructed by author GW).
The integrated results for the two authors were nearly identical; the following
discussion and figures refer to the results related to JXP.

Figure~\ref{fig:mockdz} summarizes the completeness and several biases
uncovered by our analysis.  In the top panel, all cases where a mock
$\mtll \ge 2$ LLS exists with $\mzabs > 3.2$ and an LLS was `observed'
are presented; these correspond to $> 80\%$ of the cases.  
Specifically, we plot the offset $\delta z$
between the true and
observed LLS absorption redshifts as a function of the \sna\ of the
spectrum.  
We find excellent agreement (small $\delta z$), nearly independent
of the \sna\ of the data.
There are, however, a number of cases with $\delta z > 0.01$, primarily
related to the blending of absorption lines (see below).

The lower panel in Figure~\ref{fig:mockdz} presents those cases
that are false negatives (diamonds) and false positives (triangles).
The latter are very rare.
The former, however, did occur quite often in our analysis and can
be divided into two classes:  
(i) cases with $|\delta z| < 0.05$ which are proper false negatives, i.e.\
true $\mtll \ge 2$ LLS that were modeled as a pLLS;
(ii) cases with $\delta z > 0.1$ which are sightlines where a
higher $z$ pLLS precluded the detection of a lower $z$ LLS.
The majority of the latter cases 
are due to bona-fide pLLS at higher $z$ which greatly diminish
the S/N of the spectra at shorter wavelengths and `obscure'
the presence of a $\mtll \ge 2$ LLS.
Almost none of these cases, however, satisfy the selection
criteria established in $\S$~\ref{sec:path};  either the
data have too low \sna\ or the absorption redshift of the LLS
gives $\mzem - \mzabs \ge 0.4$.  

The first class of false negatives, meanwhile,
are almost exclusively associated with spectra
that had systematically low estimates for the sky background.
In these cases, a $\mtll \ge 2$ LLS has an apparent flux 
at $\lambda < \mlll$ and therefore was modeled as a pLLS.
This is the dominant effect of a sky subtraction bias.
Although our mock spectra had even numbers of over and under-subtracted
sky backgrounds, only the former
are relatively easy to identify (large
regions of spectra are significantly negative) and ignore.
The net effect of a random sky subtraction error is a 
systematic underestimate in the incidence of LLS.
We stress, however, that the magnitude and frequency of poor
sky subtraction in the mock spectra was intentionally elevated
so that we could explore these effects.  The incidence of
such effects within the SDSS spectra is much lower, an
assertion supported by our follow-up spectra with Keck/LRIS
(Appendix~\ref{appx:keck}).
Therefore, we are confident that the sky subtraction bias 
gives rise to a less than $10\%$ systematic error for \lzt.

\subsection{The pLLS Bias}
\label{sec:pLLS}

Originally, we intended to perform a search for LLSs in spectra
with \sna=1 to maximize the pathlength of the survey 
(a nearly 4$\times$ increase over \sna=2).  Our
tests with mock spectra and follow-up observations with 
Keck/LRIS (Appendix~\ref{appx:keck}) indicated that we 
could robustly identify $\mtll \ge 2$ LLS in such data.
We also noted, however, that many of the spectra showed
pLLS candidates which reduced the \sna\ to below 1 and made the
search for $\mtll \ge 2$ LLS much more challenging.  Our
response was to redefine the search path 
by attenuating the absorbed continuum due to
any identified pLLS candidates and then reapply
the \sna\ criterion for the remaining $z<z_{\rm pLLS}$
spectral range.  With this approach, the search path was
frequently terminated by the presence of a pLLS candidate.

In principle, this modification should provide an unbiased
search for $\mtll \ge 2$ LLS.  Our trials with mock spectra,
however, revealed an insidious bias associated with this
redefinition of the search path.  Specifically,  it is very
difficult to identify pLLSs when they occur at a small
redshift offset ($\delta z \lesssim 0.1$) redward
from a $\mtll \ge 2$ LLS.
In these cases, instead of the search being terminated at
the redshift of the pLLS such that the lower $z$ LLS is not
included in the survey, it is 
the pLLS that is ignored and the LLS is included within the
statistical sample.   Furthermore, the redshift
of the recovered LLS is biased to a higher value which 
reduces the survey pathlength by a small but non-negligible
quantity.  Together, these two effects lead to an overestimate
of \lzt\ by values ranging from $30-50$\%.  
Furthermore, we find that 
it is very difficult to precisely estimate the magnitude of this
systematic bias for it depends sensitively on \lzt, the 
incidence of pLLS, and the quasar \zem\ distribution.
In our opinion, one cannot robustly correct for 
this systematic pLLS bias and we caution against performing
any analysis that would be subject to it.
For these reasons, we {\it ignore} pLLS when defining
the survey path based on a signal to noise criterion 
and performing the search for $\mtll \ge 2$ LLS.
With this approach, one must have sufficient \sna\ to identify
LLS even when one or more pLLS modify the absorbed continuum.
This last point motivated our decision to restrict the survey
to spectra with \sna~$\ge 2$.

\subsection{The Blending Bias}
\label{sec:blending}

As noted in the previous sub-section, LLS and pLLS
that lie close to one another in redshift are very difficult
to distinguish as individual systems.
This is even true in the limit where one has spectra with
exquisite S/N and resolution when $\delta z < 0.1$ (or less in the
case of high resolution echelle observations).
With SDSS spectra, the limited information provided by the Lyman
limit and the strongest Lyman series lines is insufficient
to robustly distinguish multiple LLSs from a single system.
This leads to a ``blending bias'' that manifests itself in several
ways.

First, the blending bias increases the number of LLSs
observed because pairs of pLLSs blend together to give a single
system with $\mtll \ge 2$.
Second, the absorption redshifts of the LLSs are shifted to higher
redshifts because one generally adopts \zlls\ from the higher
of the pair of systems.  
This leads to a smaller survey path and possibly a
higher inferred incidence of LLSs. 
More importantly (see below), many LLSs are shifted into the
proximate region of the quasar. 
This causes an underestimate of \lzt\
for intervening LLSs and an overestimate of PLLSs.

We explored the quantitative effects of the blending bias with
the following analysis.  We constructed a set of mock absorption
lines for each quasar in the statistical survey (Table~\ref{tab:survey})
with an incidence set to match our measurements ($\S$~\ref{sec:loz}).
Specifically, we adopted an \nhi\ frequency distribution

\begin{equation}
f(\mnhi,z) = C \mnhi^\beta \, \ltp \frac{1+z}{1+z_*} \rtp^\gamma
\label{eqn:fnmock}
\end{equation}
with $\beta = -1.3$ for $10^{16.5} \cm{-2} \le \mnhi \le 10^{19.5} \cm{-2}$
and $\beta = -2$ for $\mnhi \ge 10^{19.5}$, $z_* = 3.7$,
$\gamma = 5.1$ and $C=1.244 \sci{5}$.
From the mock absorber list we identified all LLSs with $\mnhi \ge 10^{17.5} \cm{-2}$.
This formed the control sample.
Then, we blended together all systems with $|\delta z| \le \delta z_j$
where $\delta z_j = [0.01, 0.05, 0.1, 0.2]$ and reidentified systems
satisfying $\mnhi \ge 10^{17.5} \cm{-2}$.  When blending two or more 
systems together, we set \zabs\ to the maximum of all the lines.
Finally, we calculated the incidence of LLSs using the survey
path and LLSs for each $\delta z_j$.

\begin{figure}
\centering
\includegraphics[width=3.5in]{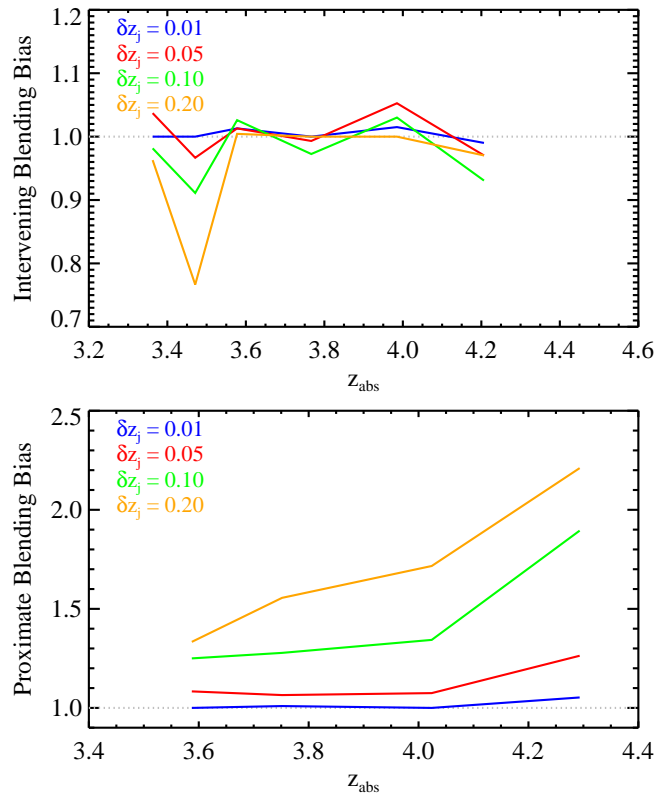}
\caption{Blending bias (enhancement/decrement of \lzt\ relative
to no bias) for mock absorption line systems for 
(upper) intervening $\mtll \ge 2$ LLSs and
(lower) proximate $\mtll \ge 2$ LLSs.
The absorption line statistics were set to roughly match the observed
incidence. 
Even for blending with redshifts $\delta z_j = 0.2$, the bias
for intervening systems is relatively small.
In contrast, the blending bias systematically elevates \lzt\ for PLLSs,
especially at $z>4$.
}
\label{fig:bbias}
\end{figure}

Figure~\ref{fig:bbias}a presents the results of the blending bias in
terms of the enhancement/decrement of the incidence of $\mtll \ge 2$ LLS
relative to the perfect model ($\delta z_j = 0$).
For $\delta z_j < 0.1$, there is only a small and ignoreable effect.
For $\delta z_j \ge 0.1$, however, we predict a systematic underestimate
for the incidence of intervening LLSs, especially at $z>4$ where
the absolute incidence is highest.  
This deficit in \lzt\ runs contrary to expectation and is entirely
due to the redshift bias where the blended LLSs end up with \zlls\
within the proximate region.  
In turn, we predict a systematic over-estimate of \lzt\ 
for PLLSs (Figure~\ref{fig:bbias}b).

Our experiments with mock spectra indicate that we commonly blend
together systems with $\delta z_j = 0.05$ to 0.1, with a weak dependence
on redshift or S/N.  
We conclude, therefore, that the effects of the blending bias
on our SDSS analysis are $< 10\%$ for measurements of \lzt\
for intervening LLSs.
This is below the current level of the statistical
error and we will ignore it in the presentation of the results and
discussion.  For the PLLSs, the effect for $\delta z_j = 0.1$ 
ranges from $20-90\%$ and we will report our results on \lzt\ 
for these absorbers as upper limits, especially for $\mzabs > 4$.

Before closing this section, we stress that the blending bias
affects all previous and future LLS surveys.  In particular,
we caution that the incidence of LLSs cannot trivially be
used to constrain the \ion{H}{1} frequency distribution \fnhi,
because the latter assumes that every absorption system is
a unique, identifiable line.  For the LLSs in particular (but
this is also true for the \lya\ forest), blending smears these
lines over non-negligible redshift intervals ($\delta z \approx 0.1$)
and this affect must be considered when comparing against 
theoretical line densities.

\begin{deluxetable*}{lccccccccc}
\tablewidth{0pc}
\tablecaption{LLS STATISTICS\label{tab:llssumm}}
\tabletypesize{\footnotesize}
\tablehead{\colhead{z} & \colhead{$\Delta X$} & \colhead{$\Delta z^a$} & \colhead{$m_{\rm LLS}^{c}$} & \colhead{$\bar z^a$} & 
\colhead{\lzt$^{d}$} &
\colhead{\lzt$^{d}$} &
\colhead{\ltlls$^d$} & \colhead{$\Delta r^e_{\rm LLS}$} \\
&&&&&(\sna=2)&(\sna=3)&(\sna=2)& (Mpc) }
\startdata
\cutinhead{Full Sample}
$\lbrack$3.30,4.40]&  366.5&   93.8& 192&3.68&$ 2.05^{+ 0.15}_{- 0.16}$&$ 2.11^{+ 0.22}_{- 0.25}$&$ 0.52^{+ 0.04}_{- 0.04}$&$ 78$\\
\cutinhead{Redshift Intervals}
$\lbrack$3.30,3.40]&   25.7&    6.8&   9&3.35&$ 1.31^{+ 0.43}_{- 0.60}$&$ 2.28^{+ 0.84}_{- 1.23}$&$ 0.35^{+ 0.11}_{- 0.16}$&$147$\\
$\lbrack$3.40,3.50]&   49.2&   12.9&  28&3.47&$ 2.17^{+ 0.40}_{- 0.49}$&$ 2.03^{+ 0.55}_{- 0.73}$&$ 0.57^{+ 0.11}_{- 0.13}$&$ 83$\\
$\lbrack$3.50,3.65]&  111.7&   29.0&  46&3.58&$ 1.59^{+ 0.23}_{- 0.27}$&$ 1.44^{+ 0.33}_{- 0.41}$&$ 0.41^{+ 0.06}_{- 0.07}$&$107$\\
$\lbrack$3.65,3.90]&  109.7&   27.9&  57&3.74&$ 2.05^{+ 0.27}_{- 0.31}$&$ 2.37^{+ 0.44}_{- 0.54}$&$ 0.52^{+ 0.07}_{- 0.08}$&$ 76$\\
$\lbrack$3.90,4.10]&   41.9&   10.4&  30&3.97&$ 2.89^{+ 0.52}_{- 0.63}$&$ 2.45^{+ 0.76}_{- 1.04}$&$ 0.72^{+ 0.13}_{- 0.16}$&$ 48$\\
$\lbrack$4.10,4.40]&   28.3&    6.8&  22&4.23&$ 3.22^{+ 0.68}_{- 0.84}$&$ 3.44^{+ 1.02}_{- 1.38}$&$ 0.78^{+ 0.16}_{- 0.20}$&$ 38$\\
\enddata
\tablecomments{Unless specified otherwise, all quantities refer to the \sna=2 threshold.  The cosmology assumed has $\Omega_\Lambda = 0.7, \Omega_m = 0.3$, and $H_0 = 72 \mkms \rm Mpc^{-1}$.}
\tablenotetext{a}{Total redshift survey path for the \sna=2 criterion.}
\tablenotetext{b}{Number of $\mtll \ge 2$ LLS discovered in the survey path.}
\tablenotetext{c}{Median absorption redshift of the LLS for the \sna=2 threshold.}
\tablenotetext{d}{{L}ine densities of LLS with $\tau \ge 2$ per redshift $dz$ or absorption length $dX$.  Often written as $dn/dz$ (or $dN/dX$) in the literature.}
\tablenotetext{e}{Average proper distance between LLS with $\tau \ge 2$.}
 
\end{deluxetable*}

\subsection{Continuum Uncertainty}
\label{sec:conti}

An important systematic uncertainty in our analysis is the placement
of the absorbed quasar continuum.  As described in $\S$~\ref{sec:continuum},
the continuum for each quasar spectrum was determined from an automated fit of a
template model to the data.  Each continuum was then reviewed
by two authors (JXP, JMO) and frequently scalled up/down by $5-10\%$.
For our LLS survey, modifications to the continuum primarily 
modify the survey path;  the estimates of \tll\ are only affected if
$\mtll \lesssim 2$.  

To test the sensitivity of our results to continuum placement, we reanalyzed
the data after scaling each continuum up/down by 5 and 10\%.
We also considered a scenario where the scaling was random between $\pm 10\%$.
We find that the estimates on the incidence of LLSs varies by $5-10\%$ as
the continua are modified.  The effect, while due to systematic changes in the
continuum, is not systematic.  That is, a systematic increase in the continuum
does not systematically increase/decrease the incidence of LLSs at all redshifts.
Therefore, we conclude that continuum placement errors yield a
random, non-negligible ($\approx 5-10\%$) uncertainty in the final results. 

\subsection{Summary}
\label{sec:mocksumm}

We have conducted an assessment of the systematic uncertainty
related to surveying LLSs using mock spectra with idealized
\lya\ forest absorption yet realistic spectral characteristics
(noise, resolution).  Our analysis revealed an insidious bias associated
with pLLSs that is best minimized by restricting the analysis to 
data with \sna$\ge 2$.  We identified an unavoidable bias related
to the blending of LLS and pLLS that implies a 
a $\approx 10\%$ uncertainty in the measured incidence of LLSs.
This bias becomes even more significant 
at $z>4$ when the incidence of LLSs exceeds 3 per unit $\Delta z$.
Finally, we found that continuum placement errors yield a
random, non-negligible ($\approx 5-10\%$) uncertainty. 
Although higher S/N and spectral resolution will reduce some
of these effects, we conclude that it will be difficult to avoid
a systematic error of $10-20\%$ using the standard approaches
to surveying LLSs.  We believe that future programs will require
new techniques if higher precision measurements are desirable.

\section{Results}
\label{sec:results}

In this section, we present the principal results of our survey.  We
defer extended discussion of previous work and the implications
of our analysis to the following section.  Systematic biases and
uncertainies in these results were discussed in the previous section
and are summarized in $\S$~\ref{sec:mocksumm}.

\subsection{\lzt: The Incidence of 
Intervening $\mtll \ge 2$ LLSs per Redshift Interval}
\label{sec:loz}

An LLS survey, by its nature, provides only a single
observable quantity:
the incidence of LLSs per redshift interval \lzt.
This quantity is independent of any assumed cosmology
and consequently has limited physical meaning.  Nevertheless,
it is the proper starting point for describing
our results.

Following standard
practice, we estimate \lzt\ from the 
ratio of $(N_{\rm LLS})$ the number of LLSs 
detected in a redshift interval to 
(\ztot) the total search path
for that redshift interval:

\begin{figure*}
\centering
\includegraphics[height=6.5in,angle=90]{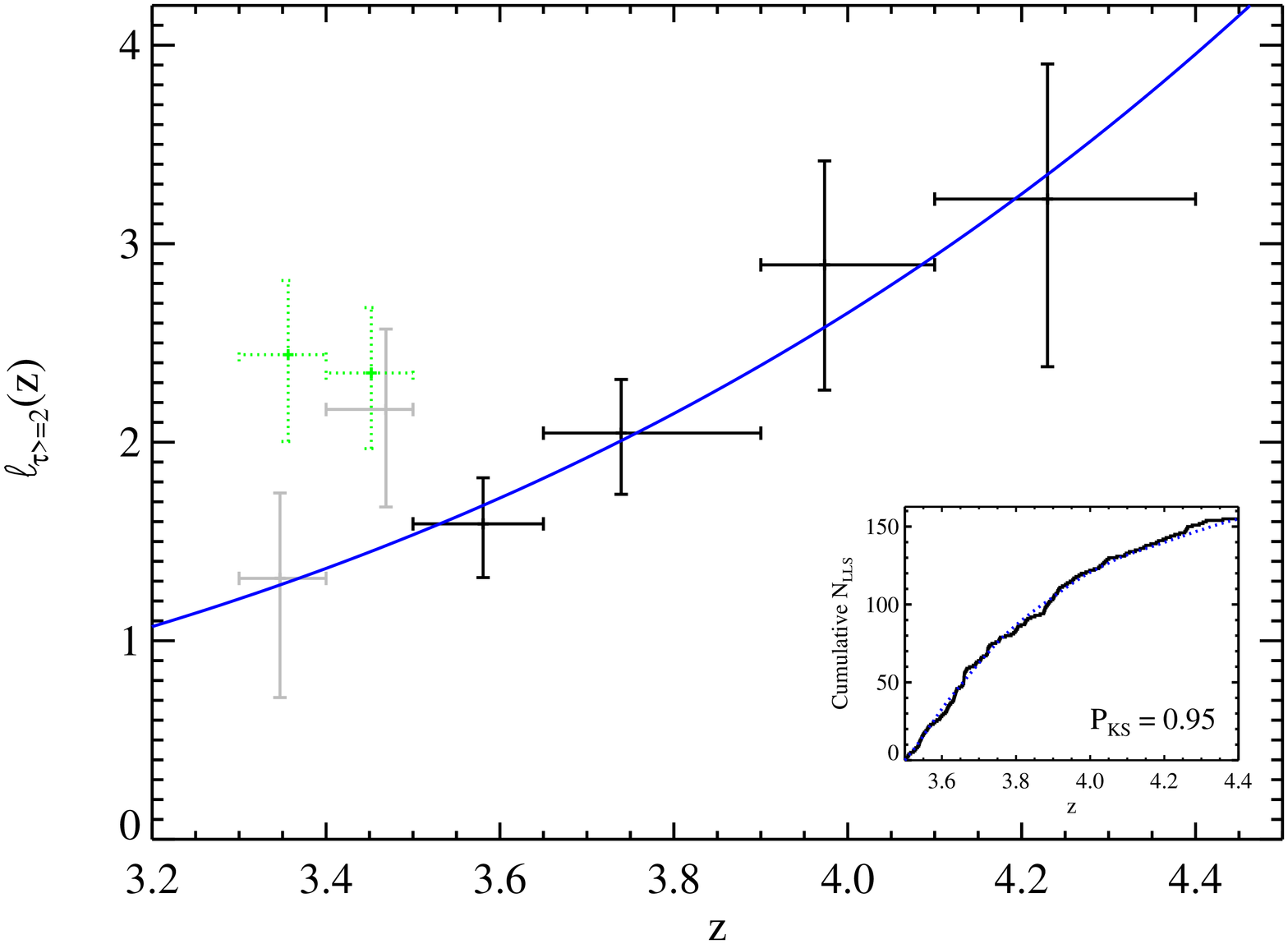}
\caption{Incidence of intervening LLSs with $\mtll \ge 2$ \lzt\
as a function of redshift (solid points).  
Only the darker points were included in a maximum likelihood
analysis to determine 
best-fit power-law (blue curve):
$\mlzt = C_{\rm LLS} [(1+z)/(1+z_*)]^\mglls$,
with $z_* \equiv 3.7$, $C_{\rm LLS} = \clls$, and
$\mglls = \alls$ (68\%\ c.l.).
The dotted points, meanwhile, show
\lzt\ when one includes quasars with $3.4 \le \mzem \le 3.6$.
These measurements are significantly biased to higher values by the
SDSS quasar-targeting criteria (PWO09).
In the subpanel, 
the black (solid) curve shows the cumulative number of
$\mtll \ge 2$ LLSs detected in our survey of the SDSS-DR7
database adopting the \sna=2 threshold.
The blue (dotted) curve shows the predicted number of LLSs assuming
the best-fit power-law from Figure~\ref{fig:loz_lls} and adopting the
$g(z)$ function from Figure~\ref{fig:goz}.  
A one-sided KS-test does not rule out the null hypothesis that
the model distribution is statistically different from the observations.
}
\label{fig:loz_lls}
\end{figure*}

\begin{equation}
\mlzt = \frac{N_{\rm LLS}}{\mztot}  \perd
\label{eqn:loz}
\end{equation}
The statistical error in \lzt\ from this estimator is assumed to be dominated
by the Poisson uncertainty in $N_{\rm LLS}$.  
We have discussed a range of possible systematic uncertainties with
this estimator in the previous section.
Figure~\ref{fig:loz_lls}
presents the values of \lzt\ for the \sna=2 criterion in
a set of arbitrary redshift intervals chosen to give $m_{\rm LLS} \gtrsim 30$
systems per bin. 
Table~\ref{tab:llssumm} lists these values for \sna\ thresholds
of 2 and 3; there is no obvious dependence with this threshold.

Figure~\ref{fig:loz_lls} reveals that the incidence of $\mtll \ge 2$ LLS
increases monotonically for $z>3.5$.
Following previous work, we have modeled the redshift
evolution in \lzt\ as a power-law with the functional form:

\begin{equation}
\mlzt = C_{\rm LLS} \ltk \frac{1+z}{1+z_*} \rtk^{\gamma_{\rm LLS}}
\label{eqn:powlaw}
\end{equation}
setting $z_* \equiv 3.7$.
Using standard maximum likelihood techniques
\citep[e.g.][]{storrie94}, we find best-fit values 
to the data at $z \ge 3.5$ of
$C_{\rm LLS} = \clls$ and 
$\mglls = \alls$ (68\%\ c.l.).  The best-fit model
is overplotted on the data in Figure~\ref{fig:loz_lls}.
The relatively large uncertainty in $\gamma_{\rm LLS}$
is due to the small redshift interval covered by our survey.
Nevertheless, we conclude at high confidence ($>95\%$)
that \lzt\ is increasing at least as steeply as $\mglls = 2$
at $z>3.5$.

The sub-panel of Figure~\ref{fig:loz_lls} compares the cumulative number of LLSs
detected in the survey against redshift both as observed (solid)
and as predicted (dotted) by the best-fit power-law model.  For the
latter, we adopt the $g(z)$ curves for \sna=2 from Figure~\ref{fig:goz}.
A one-sided Kolmogorov-Smirnov test yields a probability $P_{\rm KS} = 0.95$
that the observed distribution is drawn from the adopted power-law
expression;  the power-law model is a good
description of the observations.
We comment, however, that the best-fit slope ($\mglls = 5.2$)
is considerably steeper than most previous estimates for the LLSs
at this redshift (see $\S$~\ref{sec:lit}) and also steeper
than the redshift evolution measured for the \lya\ forest
and damped \lya\ systems \citep{kim+01,phh08}.
It is our expectation that \glls\ is likely lower than the central
value of our analysis.  This
assertion will be tested with future observations at $z<3$ and $z>4.5$.


\begin{figure*}
\centering
\includegraphics[height=6.5in,angle=90]{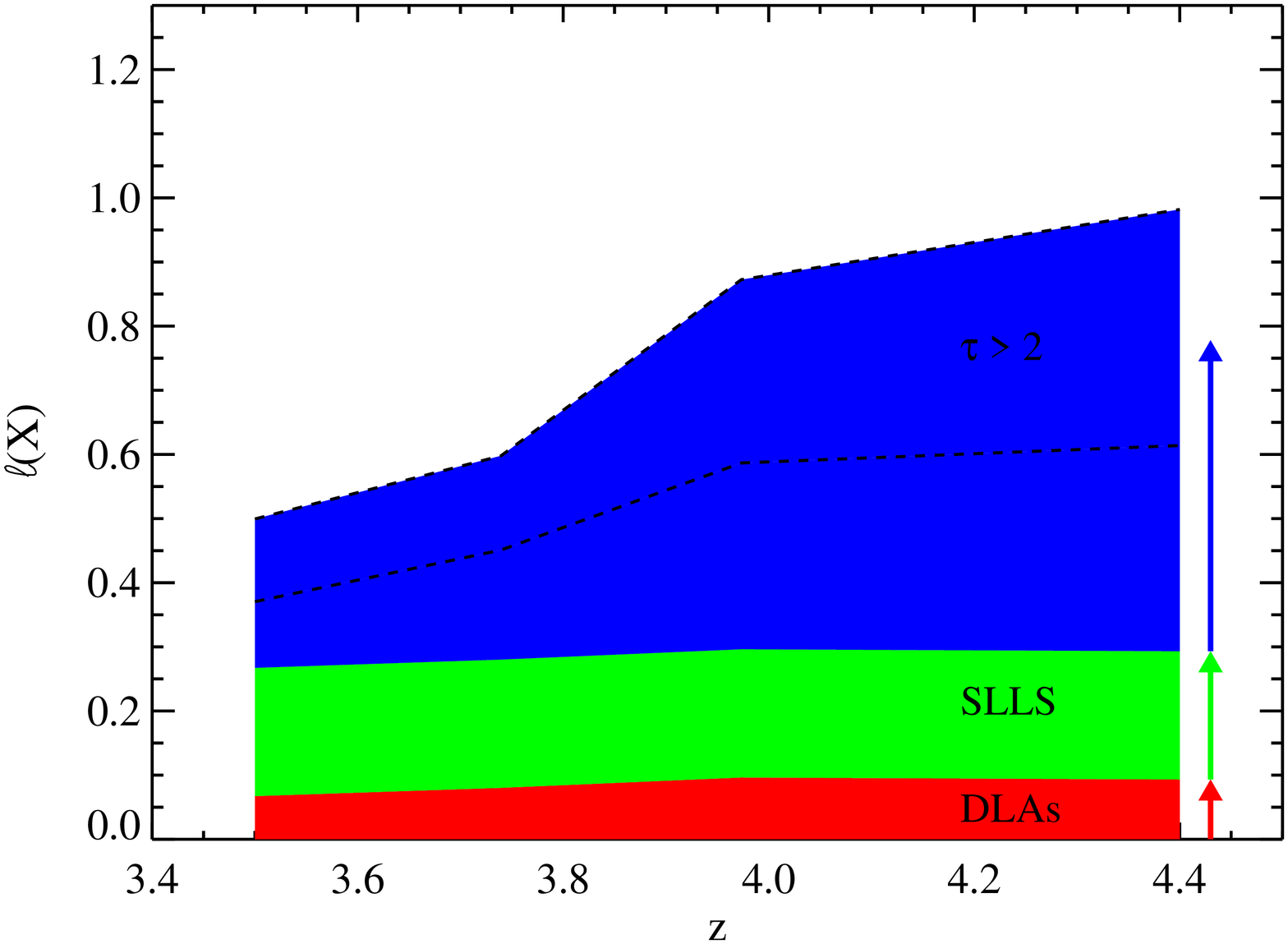}
\caption{
Incidence of $\mtll \ge 2$ LLSs as a function of absorption path $X$
as a function of redshift, indicated by the dashed lines. 
We measure a decrease of $\approx 2$ in \ltlls\ from $z=4.4$ to 3.5.
The widths of the
two lower bands (blue, red) indicate estimates of $\ell(X)$ for
DLAs and SLLSs \citep{opb+07,pw09}; see the text for details. 
These bands are plotted on top of one another to indicate their total
contribution to \ltlls\ (as indicated by the vertical arrows on the right-hand
side of the figure).  The (top) blue band, therefore, represents the 
estimate to \ltlls\ from LLSs
with $\mnhi = 10^{17.5} - 10^{19} \cm{-2}$.  Given the low absolute value
and weak evolution in $\ell(X)$ for the SLLSs and DLAs, 
the evolution in $\ell(X)$ for the LLSs is likely dominated by systems with 
$\mtll \lesssim 10$.
}
\label{fig:complox}
\end{figure*}

For $z <3.5$, Figure~\ref{fig:loz_lls} shows two evaluations
of \lzt.  The light, solid points show the \lzt\ 
values derived from our statistical quasar sample with
the restriction that $\mzem \ge 3.6$.  
These values are consistent with an extrapolation
of the best-fit power-law. 
The dotted points in the figure, meanwhile,
show the values of \lzt\ when one  also surveys
quasars with $3.4 \le \mzem \le 3.6$.  
In this case, we find systematically higher \lzt\
values which would indicate a non-physical, non-monotonic 
evolution in \lzt.  These results confirm the findings
of PWO09 that the SDSS targeting criteria for quasar 
spectroscopy biases the sample against sightlines {\it without} a LLS.
The values of \lzt\ reported in Table~\ref{tab:llssumm}, therefore,
are restricted to quasars with $\mzem \ge 3.6$.

\subsection{The Incidence of LLSs in $\Lambda$CDM}
\label{sec:lox}

If one introduces a cosmological model, the
observed incidence of $\mtll \ge 2 $ LLSs with 
redshift \lzt\ may be translated
into physical
quantities.  Consider first, \drlls,
the average distance that a photon travels
before encountering an LLS with $\mtll \ge 2$.
Specifically, we define

\begin{equation}
\mdrlls \equiv \mlzt^{-1} \frac{dr}{dz} \cmma
\end{equation}
where 
\begin{equation}
\frac{dr}{dz}  = \frac{c}{(1+z) H(z)}
\label{eqn:drdz}
\end{equation}
and
\begin{equation}
H(z) = H_0 \ltk \Omega_\Lambda + (1+z)^3 \Omega_m \rtk^{1/2} \perd
\end{equation}
With our adopted cosmology, 
we estimate that \drlls\ ranges from $\approx 100$ to $40\,\umfp$
proper distance
from $z=3.5$ to $z=4.4$ (Table~\ref{tab:llssumm}).
This is an order of magnitude or more
larger than the separation of high $z$ quasars \citep[several Mpc 
for $L_B \ge 10^{40} \rm \, erg \, s^{-1} \, Hz^{-1}$;][]{flz+09}.

\begin{figure*}
\centering
\includegraphics[height=6.5in,angle=90]{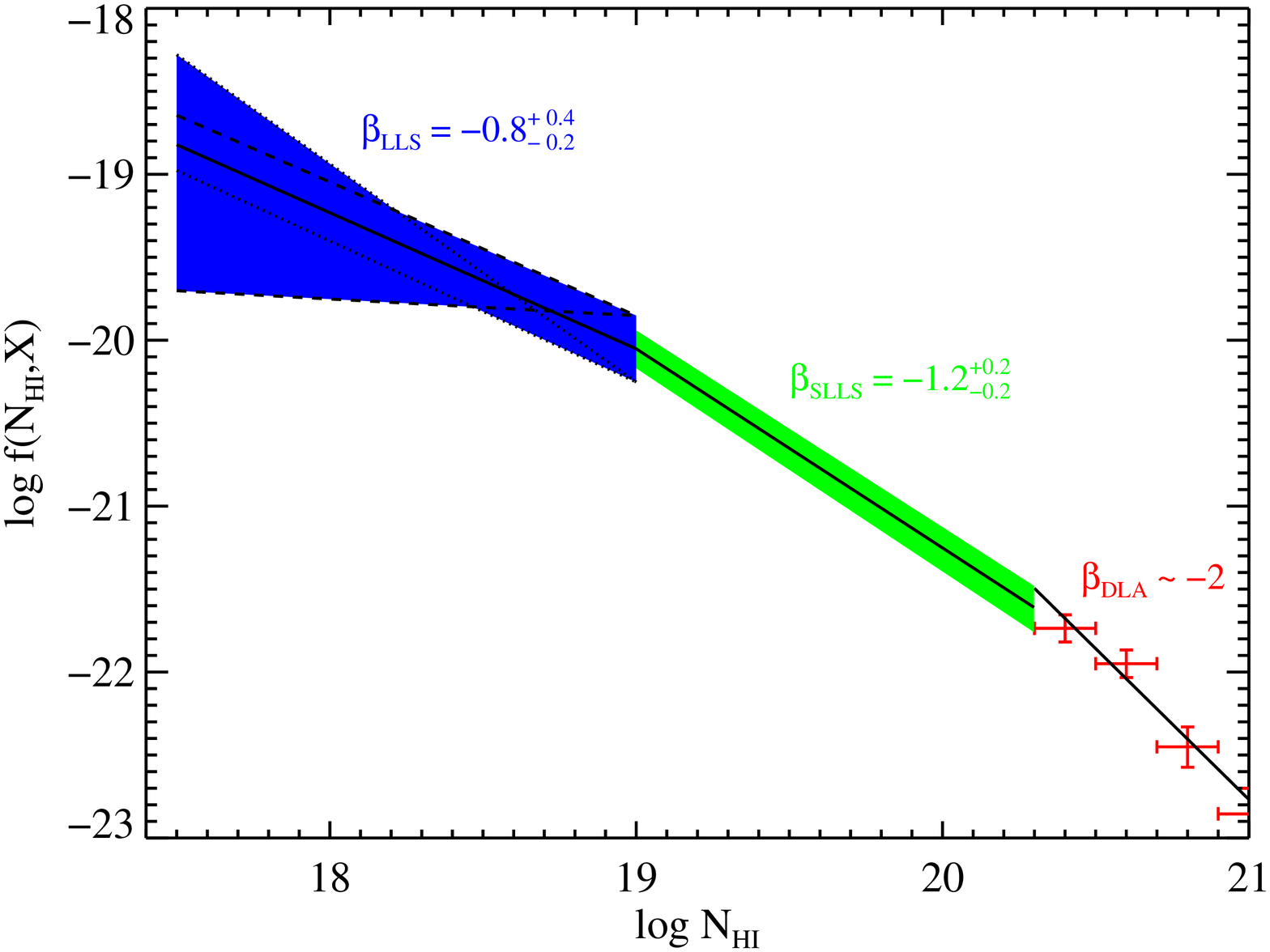}
\caption{The \nhi\ frequency distribution \fnhi\ observed for the 
SLLSs \citep[green, $\mnhi = 10^{19} - 10^{20.3} \cm{-2}$;][]{opb+07}
and DLAs \citep[red, $\mnhi \ge 10^{20.3} \cm{-2}$;][]{pw09}.
The blue band is an estimate of \fnhi\ for LLSs having 
$\mnhi = 10^{17.5} - 10^{19} \cm{-2}$ under the assumptions
of a power-law form ($f(\mnhi,X) \propto \mnhi^{\beta_{\rm LLS}}$)
and constrained by the observed incidence of SLLSs and
$\tau \ge 2$ LLS (this paper).
We find $\mbtlls = -0.8^{+0.4}_{-0.2}$ (68\%\ c.l.) for
conservative estimates on the value of \fnhi\ at $10^{19} \cm{-2}$
and allowing for 20\%\ uncertainty in \ltlls.
The dashed and dotted curves indicate the range of power-laws that
satisfy the observations.
}
\label{fig:fn}
\end{figure*}

An especially informative quantity for 
associating LLSs to structures in the Universe (e.g.\ galaxies, filaments)
is $\ell(X)$
the number of systems per absorption length \citep[][]{bp69},
where $\ell(X) dX = \ell(z) dz$ and 

\begin{equation}
dX = \frac{H_0}{H(z)}(1+z)^2 dz  \perd
\end{equation}
The quantity \ltlls\ is defined to remain constant if \nlls, the
comoving number density
of structures giving rise to a $\mtll \ge 2$ LLS,
times \slls, the average physical size of the structure 
remains constant, i.e.\ $\mltlls \propto \mnlls \mslls$.
Figure~\ref{fig:complox} presents the evolution of \ltlls\
for our cosmology as a function of redshift (see also
Table~\ref{tab:llssumm}).  
We observe a rise in \ltlls\ with redshift of roughly two times
over the $\approx 1\,$Gyr from $z=3.3$ to 4.4.
At 99\% confidence, we infer an increase in \ltlls\ over
this redshift interval.
This follows, of course, from the very steep
redshift evolution observed for \lzt\ ($\S$~\ref{sec:loz});
in a flat cosmology with $\Omega_{\rm m}$ on the order of $\Omega_\Lambda$,
an \lzt\ evolution steeper
than $(1+z)^{1/2}$ implies \ltlls\ is also increasing.
We conclude
that \nlls\ and/or \slls\ are increasing with redshift
at $z \approx 3.5$.  
We discuss the implications of this result in $\S$~\ref{sec:nature}.

\subsection{\fnhi\ at $z \approx 3.7$}
\label{sec:fnhi}

In this subsection, we combine our results with previous work
on the IGM to place constraints on the \ion{H}{1} frequency 
distribution, \fnhi.  We focus this analysis at a single redshift
($z=3.7$) where our observations have greatest statistical power.

\subsubsection{\lfnhi}

Although our observations and LLS analysis are insensitive
to the \ion{H}{1} column densities of the LLSs, 
they do provide an integral
constraint on the frequency distribution of \nhi, \fnhi.
We constrain the
\nhi\ frequency distribution of $\mtll \lesssim 10$
LLS per absorption length,
\lfnhi, at column densities
$\mnhi = 10^{17.5} - 10^{19} \cm{-2}$ as follows.
Previous surveys at $z>3$ 
have measured \fnhi\ for column densities 
$\mnhi \ge 10^{19} \cm{-2}$ \citep{phw05,opb+07,pw09,np+09,gp+09}.
These authors have parameterized the
distribution functions as single (SLLSs) and double (DLAs) power-laws
of the following form:

\begin{equation}
f_{\rm SLLS}(10^{19} \cm{-2} \le \mnhi < 10^{20.3} \cm{-2},X) 
  = k_{\rm SLLS} \mnhi^{\beta_{\rm SLLS}} 
\label{eqn:fslls}
\end{equation}

and

\begin{eqnarray}
&& f_{\rm DLA}(\mnhi \ge 10^{20.3} \cm{-2},X)  
 = k_{\rm DLA} \ltp \frac{\mnhi}{N_d} \rtp^{\beta_{\rm DLA}}
\nonumber \\
&&  \; {\rm where} 
\; \beta_{\rm DLA} =  
\begin{cases}
\beta_3:  \mnhi < N_d ; \quad \\ 
\beta_4:  \mnhi \geq N_d \\
\end{cases}
\label{eqn:fdla}
\end{eqnarray}
Figure~\ref{fig:fn} presents these frequency distributions.
For the SLLSs at $z=3.7$, we have taken $\beta_{\rm SLLS} = -1.2 \pm 0.2$ and
normalized the power-law by taking 

\begin{equation}
\mlslls = \intl_{10^{19}\cm{-2}}^{10^{20.3}\cm{-2}} \msfnhi \, d\mnhi = 0.2 \perd
\end{equation}
These values are consistent with the range of published measurements
at this redshift
\citep{peroux05,opb+07,gp+09}.
For the DLAs, we have evaluated \dfnhi\ from the SDSS-DR5 \citep{pw09}
over the redshift interval $z=[3.4,4.0]$, giving 
$N_d = 10^{21.75}$, $\beta_3 = -1.8$, $\beta_4 < -3$, 
and $k_{\rm DLA} = 7\sci{-25} \cm{2}$. 

We estimate \fnhi\ for the interval $\mnhi = [10^{17.5}, 10^{19}]\cm{-2}$,
which we refer to as \lfnhi,
under the following assumptions/constraints:
(i) \lfnhi\ has a power-law form 

\begin{equation}
\mlfnhi = k_{\rm LLS} \mnhi^{\beta_{\rm LLS}} \cmma
\end{equation}
and (ii) \lfnhi\ at $\mnhi=10^{19}\cm{-2}$ is consistent with the
range of values given by the SLLSs.
Specifically, we demand $\log \mvfnhi{19} = -20.05 \pm 0.2$;
(iii) we impose the integral constraint based on the
the observed incidence of $\mtll \ge 2$ LLSs:

\begin{equation}
\mltlls = \intl_{10^{17.5} \cm{-2}}^\infty \mfnhi \, d\mnhi \perd
\end{equation}
At $z \approx 3.7$, we estimate $\mltlls = 0.5 \pm 0.1$ (Figure~\ref{fig:complox}).

Overplotted on Figure~\ref{fig:fn} are the power-law frequency distributions
(shown as dashed, dotted and a solid line) that satisfy the extrema
of those constraints.
The shaded region shows the intersection of the curves and roughly 
represents the allowed region of \fnhi\ values.
We find $\mbtlls = \blls$, and derive $k_{\rm LLS} = 10^{-4.5} \cm{2}$
for the central value.  Table~\ref{tab:fn} further summarizes
these results.

Our analysis reveals that \fnhi\ becomes increasingly shallow
with decreasing \nhi. Only for the most extreme
values of our analysis, low \fnhi\ at $\mnhi = 10^{19}\cm{-2}$ and
a large \lzt\ value, do we recover $\mbtlls < -1$.
This flattening of \fnhi\ was suggested by previous authors
based on a similar analysis but with much poorer observational
constraints on \ltlls\ \citep{peroux_dla03,phw05,opb+07}.
Remarkably, our results indicate $\mbtlls > -1$
which means that the IGM has a {\it higher} total covering 
fraction per unit pathlength for 
sightlines with $\mnhi = 10^{19} \cm{-2}$ than those 
with $\mnhi = 10^{18} \cm{-2}$.

\subsubsection{Constraints from Measurements of the Mean Free Path}
\label{sec:mfp}

Traditionally, the mean free path to ionizing radiation in the IGM (\lmfp)
has been estimated from the observed incidence of LLSs \citep[e.g.][]{mm93}.
Recently, PWO09 introduced a new approach to measure \lmfp\ from
stacked quasar spectra, without any consideration of LLSs.
We can reverse the problem, therefore, and use the \lmfp\ results
to constrain properties of \fnhi.  One expects to have the greatest
sensitivity to absorption systems with $\mtll \approx 1$, i.e.\ the LLSs
and partial LLSs.

At $z=3.7$, PWO09 estimate $\mlmfp = 47 h^{-1}_{72} \, \rm Mpc$ proper distance.
This means that in the absence of an expanding universe, a packet of
1\,Ryd photons at $z=3.7$ would be attenuated by $\exp(-1)$ after traveling
\lmfp.
One can also express the mean free path as an opacity,
\kll=1/\lmfp, which can be related to the optical depth of a
1\,Ryd photon as:

\begin{equation}
\mkll = \frac{d\mteff}{dr} = \frac{d\mteff}{dz} \frac{dz}{dr}.
\label{eqn:kappa}
\end{equation}
Finally, we can relate the differential optical depth to the \ion{H}{1}
frequency distribution of absorbers:

\begin{equation}
\frac{d\mteff(z)}{dz} = \intl_{\mnmin}^\infty f(\mnhi,z)
   \lbrace 1 - \exp \ltk - \mnhi \sigma_{\rm ph}^{912} \rtk \rbrace d\mnhi
\label{eqn:teff}
\end{equation}
with $\sigma_{\rm ph}^{912}$ the photoionization cross-section evaluated
at 1\,Ryd and $dr/dz$ given by Equation~\ref{eqn:drdz}.
Although the integral should be evaluated with $\mnmin = 0$, in
practice $d\mteff$ is insensitive to the minimum \nhi\ column density 
for any value $\mnmin \le 10^{12} \cm{-2}$.

\begin{figure}
\centering
\includegraphics[height=3.5in,angle=90]{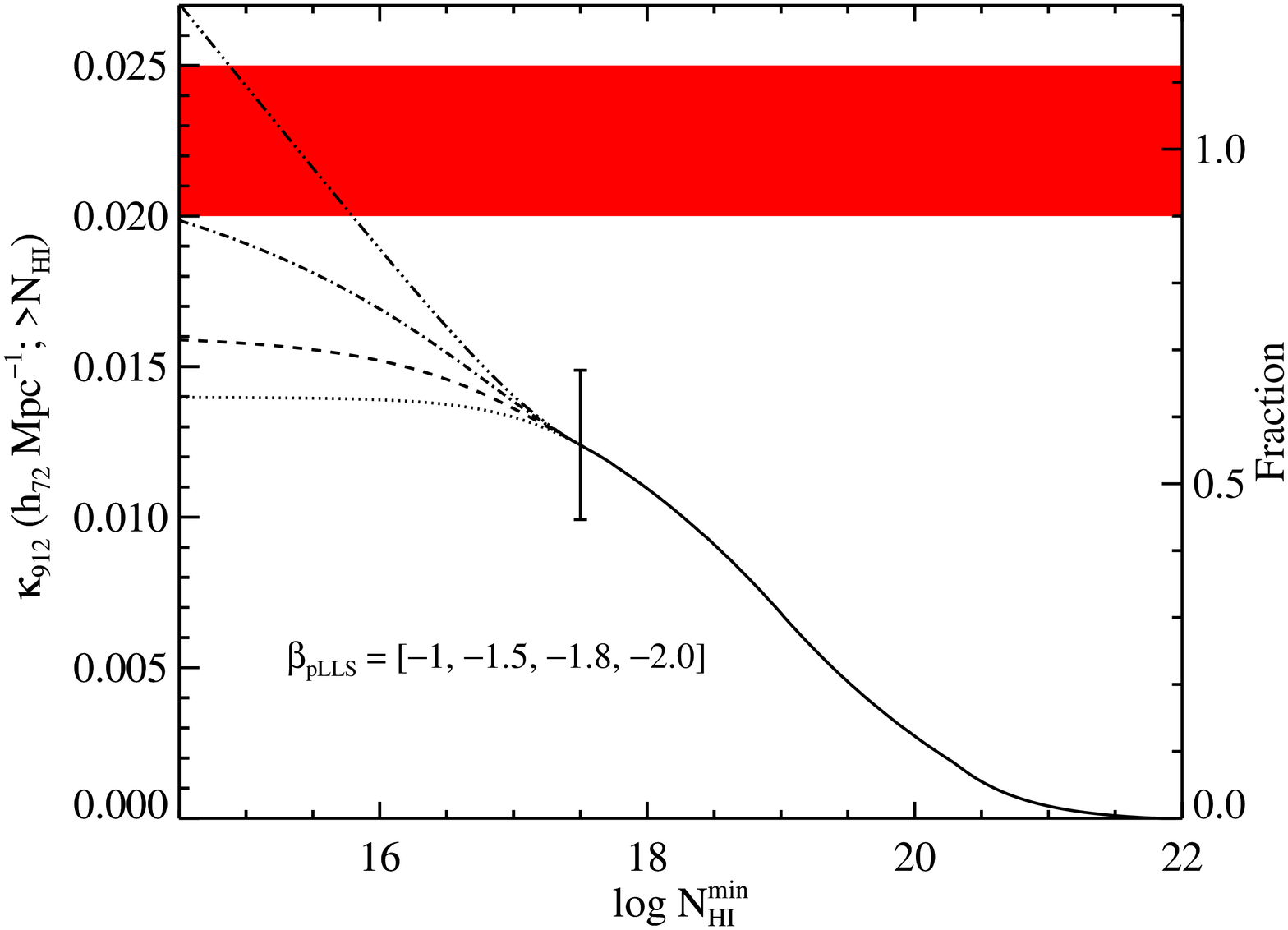}
\caption{This figure shows the opacity at the Lyman limit \kll\ contributed
by absorbers with $\mnhi \ge \mnmin$. For $\mnmin \ge 10^{17.5} \cm{-2}$
(solid curve), which corresponds to our LLS survey, we have adopted
the estimate of \fnhi\ from Figure~\ref{fig:fn} in the calculation.
We estimate a 20\%\ uncertainty in the contribution of LLSs to
\kll, as shown in the figure.  For $\mnmin < 10^{17.5} \cm{-2}$,
we assume \fnhi\ follows a simple power-law with exponent \bplls,
and show a series of extrapolations (dash and dotted curves).
The solid (red) horizontal band centered at 
$\mkll = 0.0225 \, h_{72} \, \rm Mpc^{-1}$
indicates the measurement at $z \approx 3.7$ by PWO09.
These results imply that LLSs contribute $\approx 55\%$ of the 
mean free path ($\approx 33\%$ for systems with $\mtll \gg 1$)
and that \bplls\ must be steeper than $\approx -1.5$
to explain all of these observations.
}
\label{fig:kappa}
\end{figure}

Figure~\ref{fig:kappa} shows the \kll\ value at $z=3.7$ from PWO09
as a horizontal band that illustrates the $1\sigma$ error interval.
The solid curve, meanwhile, corresponds to the evaluation of
Equation~\ref{eqn:kappa} using our best estimation of \fnhi\ 
(Figure~\ref{fig:fn}) as a cumulative function of \nmin.
At the limiting \nhi\ value of our LLS survey ($10^{17.5} \cm{-2}$), 
we estimate that $\approx 55\%$ of the opacity to ionizing radiation is
contributed by $\mtll \ge 2$ LLSs.  
The uncertainty in the results is roughly proportional to the
uncertainty in \lzt, i.e. $\approx 20\%$ as indicated by the error
bars on the figure.
It is notable that $\approx 1/3$ of the contribution to \kll\
is from very optically thick absorbers ($\mtll \gg 1$), 
i.e.\ the SLLSs and DLAs.

\begin{figure*}
\centering
\includegraphics[height=6.5in,angle=90]{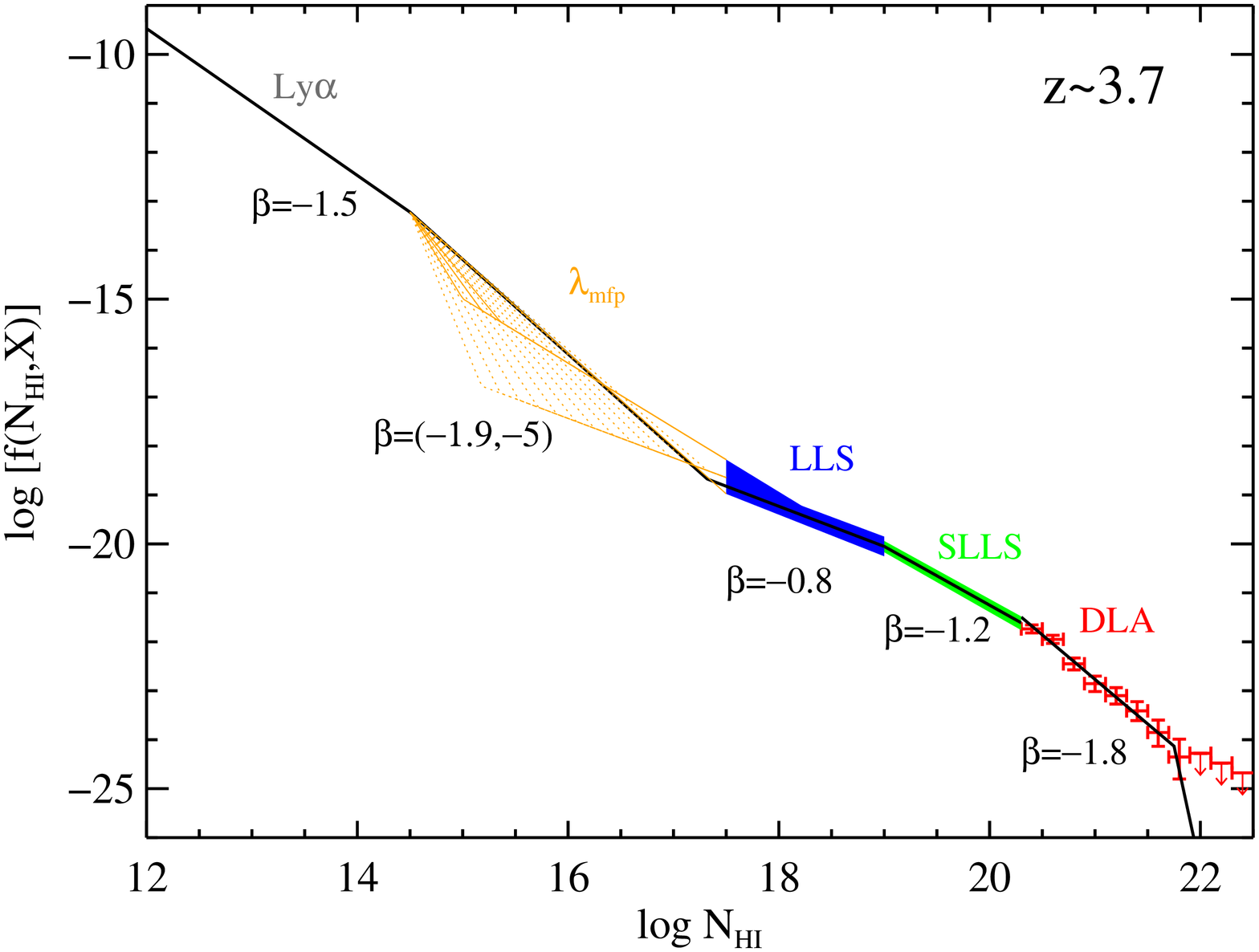}
\caption{The solid black curve shows our estimation of
\fnhi\ at $z\approx 3.7$ as a series of six power-laws that intersect at
$\mnhi = [10^{14.5}, \mnplls, 10^{19.0}, 10^{20.3}, 10^{21.75}] \cm{-2}$,
where \nplls\ is constrained to lie between $\mnhi = 10^{15} - 10^{17.5} \cm{-2}$.
The observational constraints for \fnhi\ at $\mnhi \ge 10^{19} \cm{-2}$
are as in Figure~\ref{fig:fn}.  The results here also include 
constraints from the observed mean free path (PWO09) and the effective
\lya\ opacity of the \lya\ forest \citep{fpl+08}.
The orange curves
show the regions of \fnhi\ that reproduce the mean free path
measurement and also connect to the \lya\ forest at 
$\mnhi = 10^{14.5} \cm{-2}$. 
}
\label{fig:powerfn}
\end{figure*}

It is also evident from Figure~\ref{fig:kappa} that systems with 
$\mtll \le 2$ must contribute to \kll.
For $\mnmin < 10^{17.5} \cm{-2}$, we continue the calculation
by assuming that \fnhi\ follows a power-law

\begin{equation}
f_{\rm pLLS}(\mnhi < 10^{17.5} \cm{-2}, X) = 
  k_{\rm pLLS} \mnhi^{\beta_{\rm pLLS}} 
\label{eqn:fplls}
\end{equation}
constrained to match \lfnhi\ at $\mnhi = 10^{17.5} \cm{-2}$.
We find that models with $\mbplls \ge -1.8$ cannot reproduce
the \lmfp\ results.  In fact, the data favor $\mbplls \approx -2$
i.e.\ a much steeper power-law than inferred for the LLS and also
than that commonly observed for the \lya\ forest.  These conclusions
depend rather insensitively on our
estimate of \ltlls; slopes only as shallow as $-1.7$ are allowed
if we adopt our highest estimates for \ltlls.

Thus far, these
inferences on \fnhi\ for absorption systems with 
$\mtll < 2$ have ignored observations of the \lya\ forest.
By including these data, we provide further constraints
on \fnhi\ for $\mnhi = 10^{15} \cm{-2}$ to $10^{19} \cm{-2}$.
To derive these constraints, however, we must adopt a functional
form for \fnhi.  
Absent a physical model, we take an empirical approach.
We express \fnhi\ as a series of six power-laws that
intersect at 
$\mnhi = [10^{14.5}, \mnplls, 10^{19.0}, 10^{20.3}, 10^{21.75}] \cm{-2}$,
where \nplls\ is constrained to lie between $\mnhi = 10^{15} - 10^{17.5} \cm{-2}$.
Other than a small `kink' at $\mnhi = 10^{20.3} \cm{-2}$, the power-laws
are required to match at each intersection point.

\begin{deluxetable}{cccccc}
\tablewidth{0pc}
\tablecaption{SUMMARY OF \fnhi\ RESULTS \label{tab:fn}}
\tabletypesize{\footnotesize}
\tablehead{\colhead{\lzlls$^a$} & \colhead{$\log f_{19}^b$}
& \colhead{\btlls} & \colhead{$\log k_{\rm LLS}$} & \colhead{log \nplls$^c$}
& \colhead{\bplls}}
\startdata
\cutinhead{Preferred Values}
0.23&$-20.05$&$-0.8$&$ -4.5$&17.3&$-1.9$\\
\cutinhead{Conservative Range of Allowed Values}
0.23&$-20.05$&$-0.8$&$ -4.5$&17.1&$-2.0$\\
&&&&17.3&$-1.9$\\
&&&&17.5&$-1.9$\\
0.15&$-20.25$&$-0.9$&$ -4.1$&17.5&$-1.9$\\
0.35&$-20.25$&$-1.3$&$  4.7$&15.0&$-3.5$\\
&&&&15.2&$-3.0$\\
&&&&15.4&$-2.6$\\
0.15&$-19.85$&$-0.1$&$-18.0$&\dots & \dots\\
0.35&$-19.85$&$-0.8$&$ -4.6$&15.2&$-5.2$\\
&&&&15.4&$-4.3$\\
&&&&15.5&$-3.7$\\
&&&&15.7&$-3.3$\\
&&&&15.9&$-3.0$\\
&&&&16.1&$-2.7$\\
&&&&16.2&$-2.5$\\
&&&&16.4&$-2.4$\\
&&&&16.6&$-2.2$\\
&&&&16.8&$-2.1$\\
&&&&17.0&$-2.0$\\
&&&&17.1&$-1.9$\\
&&&&17.3&$-1.9$\\
\enddata
\tablecomments{The analysis throughout assumes that the DLAs contribute $\ell(X) = 0.09$ to \ltlls.}
\tablenotetext{a}{The incidence of $\mtll \ge 2$ LLS \ltlls, attributed to LLS with $\mnhi < 10^{19} \cm{-2}$.}
\tablenotetext{b}{The adopted value of \fnhi\ at $\mnhi = 10^{19} \cm{-2}$.}
\tablenotetext{c}{The break column density within the \lya\ forest as defined in the text.  Entries without values have power-law descriptions for the LLS that cannot satisfy the mean free path and \lya\ forest constraints.}
\end{deluxetable}

The power laws are forced to satisfy the following
observational constraints:

\begin{enumerate}
\item The power-laws for $\mnhi \ge 10^{17.5}$ are constrained
as described at the start of this sub-section.

\item For the \lya\ forest, we assume $\mfnhi \propto \mnhi^{-1.5}$
and normalize at $z=3.7$ by the
effective \lya\ optical depth \lteff\ measured by \cite{fpl+08}
assuming a $b$-value distribution
$f(b) \propto b^{-5} \exp[-b_\sigma^4/b^4]$ \citep{hr99}. 

\item The integrated opacity of the IGM at the Lyman limit is
constrained by the measurement of PWO09, i.e.\ $\mlmfp = 47 \umfp$.
\end{enumerate}
For the range of power-laws derived from our analysis of the LLS results
(Figure~\ref{fig:fn}), we show in Figure~\ref{fig:powerfn} the
range of \fnhi\ distributions that also satisfy all of 
the constraints.
We find viable models with \nplls\ values that range across the allowed
interval.  These are  
correlated with \bplls\ values ranging from $\mbplls = -1.9$ to $-5$
(Table~\ref{tab:fn}).  

The principal results of this analysis are threefold.  
First, the single power-law connecting the \lya\ forest to the SLLS
satisfies neither the LLS nor mean free path constraints.  
There is 
at least one break between $\mnhi = 10^{14.5} \cm{-2}$ and
$\mnhi = 10^{19} \cm{-2}$ where \fnhi\ steepens to $\beta < -1.8$
and then flattens to $\beta \approx -1$.  This is consistent with 
conclusions drawn from line-counting statistics of \lya\ forest
lines \citep[e.g.][]{petit93,kim02}.
Second, the added \lmfp\ and \lya\ forest constraints rule out
the lowest values of \lfnhi\ at $\mnhi \approx 10^{17.5} \cm{-2}$
that were otherwise allowed by our 
LLS results.  Specficially, the data require $\log f(\mnhi=10^{17}\cm{-2},X) \ge -19$
and the (blue) shaded region in Figure~\ref{fig:powerfn} shows the proper
allowed range for \lfnhi.
Finally, we find that the slope of \fnhi\ must steepen to
$\mbplls \le -1.8$ at columns $\mnhi \gtrsim 10^{14.5} \cm{-2}$.

\begin{deluxetable}{lcc}
\tablewidth{0pc}
\tablecaption{SDSS-DR7 PROXIMATE $\mtll \ge 2$ LLS SURVEY\label{tab:prox}}
\tabletypesize{\footnotesize}
\tablehead{\colhead{Quasar} &\colhead{\zem} & 
\colhead{\zlls}}
\startdata
J001115.23$+14$4601.8&4.967&\ldots \\
J001820.71$+14$1851.5&3.936&\ldots \\
J004219.74$-10$2009.4&3.880&\ldots \\
J004240.65$+14$1529.6&3.687&3.684\\
J010619.24$+00$4823.3&4.449&\ldots \\
J011351.96$-09$3551.0&3.668&\ldots \\
J012403.77$+00$4432.7&3.834&\ldots \\
J014049.18$-08$3942.5&3.713&3.693\\
J015048.82$+00$4126.2&3.702&\ldots \\
J015339.61$-00$1104.8&4.194&\ldots \\
J021318.98$-09$0458.3&3.794&3.797\\
J022518.35$-00$1332.2&3.628&\ldots \\
J024447.78$-08$1606.1&4.068&\ldots \\
J025518.58$+00$4847.6&3.989&\ldots \\
J031213.97$-06$2658.8&4.031&\ldots \\
J034402.85$-06$5300.6&3.957&\ldots \\
J073149.50$+28$5448.6&3.676&\ldots \\
J074154.59$+34$1252.1&3.905&\ldots \\
J074500.47$+34$1731.1&3.713&\ldots \\
J074640.16$+34$4624.7&4.010&\ldots \\
J074711.15$+27$3903.3&4.154&\ldots \\
J075006.62$+49$1834.1&3.603&\ldots \\
J075017.49$+40$5825.3&3.864&3.849\\
J075103.95$+42$4211.6&4.163&\ldots \\
J075347.41$+28$1805.2&4.031&\ldots \\
J075552.41$+13$4551.1&3.673&\ldots \\
J075732.89$+44$1424.6&4.170&\ldots \\
\enddata
\tablecomments{[The complete version of this table is in the electronic edition of the Journal.  The printed edition contains only a sample.]}
\end{deluxetable}

\subsection{The Incidence of Proximate LLSs (PLLSs)}
\label{sec:prox}

The results presented thus far all refer to intervening LLSs, i.e., systems
restricted to have \zlls\ blueward of 3000\kms\ from the quasar
emission redshift.  This restriction was imposed to isolate the
`ambient' IGM and avoid biases related to having performed
the search for bright, background quasars.
In the space surrounding a bright quasar, one predicts at least two
such biases:
(1) bright, high $z$ quasars are known to cluster strongly
\citep[$r_0 > 15 \, h^{-1} \rm Mpc$;][]{shen07}
suggesting these objects trace massive structures in the young universe.
The local environment of bright quasars, therefore, has an uncommonly
high density (at least in dark matter) which may give a higher
incidence of LLSs;
(2) the radiation field of the quasar will ionize gas to large
distances, reducing the incidence of LLSs.
For damped \lya\ systems, the first effect dominates
as one observes an enhanced rate of proximate DLAs (PDLAs)
relative to the intervening systems \citep{reb06,phh08}.

\begin{figure}
\centering
\includegraphics[height=3.5in,angle=90]{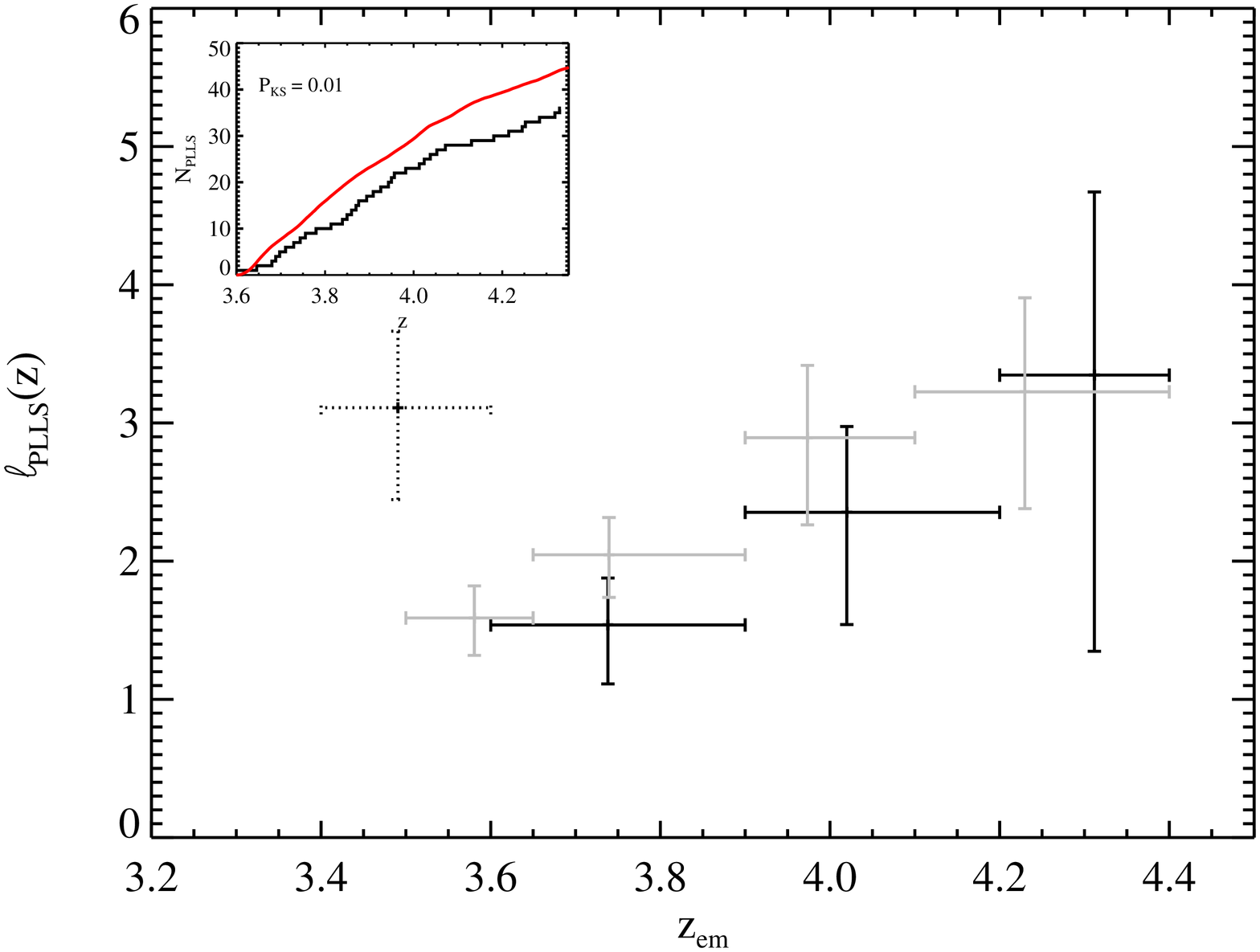}
\caption{The solid and dark points show the incidence \lplls\ of
proximate LLSs 
(PLLSs; LLSs with $\mtll \ge 2$ that 
occur within 3000~\kms\ of the background quasar) 
per unit redshift against quasar emission redshift.
These are compared against the same quantity for intervening LLSs
(gray points).  The data point at $\mzem<3.6$ 
has been dotted out because it is biased
by the SDSS targeting criteria (PWO09).  Ignoring that last
point, we find that the incidence of PLLSs roughly follows that
of intervening systems but is systematically lower by $\approx 25\%$
at $z<4$.  The inset figure shows the cumulative number of 
PLLSs observed (black curve) against the predicted number using
the power-law model for \lzt\ ($\S$~\ref{sec:loz}) and
the $g(z)$ curve for PLLSs (Figure~\ref{fig:goz}).
}
\label{fig:prox}
\end{figure}

Following the formalism presented in \cite{phh08}, we have estimated
the incidence of PLLSs in a series of redshift intervals.
First, we re-measured the quasar emission redshifts for all
systems with $\msna \ge 2$ at the Lyman limit and with 
a $\mtll \ge 2$ LLS within 5000\kms\ of \zem.
The SDSS quasar redshifts reported in the standard DR7 data release
are known to have significant systematic errors. 
Following the prescriptions described in \cite{shen07},
J. Hennawi has kindly remeasured the redshifts for all of the quasars.
Second, we analyzed the quasars whose absorbed continuum at the wavelength
of the Lyman limit corresponding to 3000\kms\ blueward of \zem\ is
twice the median-smoothed, $1\sigma$ error-array.  This establishes
the survey path.  All PLLSs identified redward of this 3000\kms\ offset
form the statistical sample (Table~\ref{tab:prox}).
The incidence, \lplls, is then estimated in arbitrary redshift
intervals assuming the same estimator for intervening 
LLSs (Equation~\ref{eqn:loz}).
These results are presented in Figure~\ref{fig:prox} and compared
against the incidence of intervening LLSs.

Ignoring the data at $z < 3.6$ (which we suspect to be biased high
by the SDSS targeting criteria; PWO09), the
incidence of PLLSs roughly tracks that of intervening LLSs but is
$\approx 25\%$ lower than the intervening systems.  The inset figure
shows the observed cumulative number of PLLSs (dark curve) versus
the predicted number (light curve) assuming the best-fit power-law
for \lzt\ of intervening LLSs.
A one-sided KS test yields only a 1\%\ probability that the
two distributions are drawn from the same parent population.
We also remind the reader that corrections for the blending bias
described in $\S$~\ref{sec:blending} will likely reduce \lplls\
further, especially at $z \sim 4$.
The principal implication is that $\mtll \ge 2$ LLSs toward
$i<20$\,mag quasars at $z>3.5$ suffer from a proximity effect,
presumably due to the ionizing radiation
field of the quasar itself.
Given the observed enhancement of strong LLSs at $z \approx \mzem$
along sightlines {\it transverse} to such quasars,
our results lend further evidence that quasar emission
at $\approx 1$\,Ryd is usually anisotropic \citep{QPQ2}.

Before concluding this section, we comment that the PLLS
analysis is subject to another systematic error.
In performing our LLS survey, we have identified and removed
all quasars with very strong associated absorption,
e.g.\ BALs.  The intent of this procedure was to remove the
signatures of absorption from gas very local 
($<1$\,kpc) to the quasar from the analysis.  It is possible, however,
that the associated absorption in some of these removed
quasars is due to a PLLS at distances $\gg 1$\,kpc and not very
local gas.
This would lead to an underestimate of \lplls.
Alternatively, we may not have identified all of the
local absorbers and therefore might have overestimated \lplls.
In either case, we caution that a systematic error of the
order of $10-20\%$ should be attributed to this effect.


\section{Discussion}
\label{sec:discuss}

\subsection{Comparisons with Previous Work}
\label{sec:lit}

Surveys for Lyman limit systems have been carried out for several
decades now \citep{tytler82,ssb,lzt91,storrie94,key_lls,peroux_dla03}.
These have been  performed primarily at optical wavelengths on 
heterogeneous quasar samples drawn from a diverse set of
survey approaches: color-selection, radio detection, slitless
spectroscopy, etc.  The authors adopted differing completeness
limits for \tll\ (ranging from 1 to 3) and used different approaches
to establishing the pathlength that establishes \lzt.  Little
attention was given to assessing systematic error, and several
of the effects described in $\S$~\ref{sec:mock} assuredly apply
to the previous works.  Perhaps not surprisingly, therefore,
most of the previous estimates of $\ell(z)$ are in disagreement
with our results.

\begin{figure}[ht]
\centering
\includegraphics[height=3.5in,angle=90]{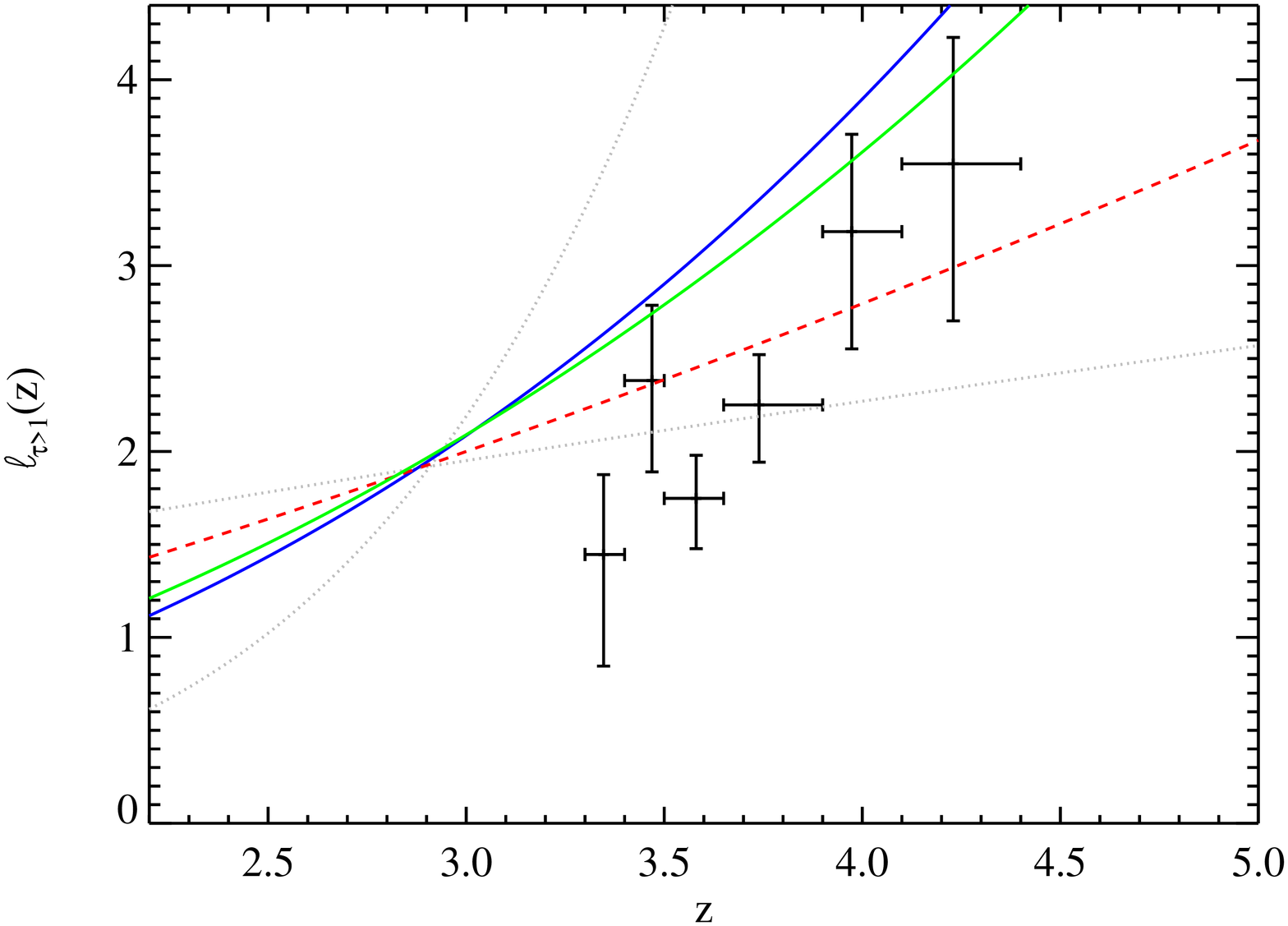}
\caption{Estimates of \lzo\
from several previous studies, parameterized as power-laws with the
parameters listed in Table~\ref{tab:param}.
The dotted (gray) lines show estimates from \cite{ssb} and \cite{lzt91}.
The dashed curve is the estimate from \cite{key_lls} and 
the solid blue and green curves are from \cite{storrie94} and \cite{peroux_dla03}
respectively.
For the values of our survey, we have increased the \lzt\ results
by 10\%\ to match the $\mtll = 1$ threshold of the previous work.
Our results indicate a significantly lower incidence of LLSs at $z<4$
than suggested by the previous estimates.
This lower incidence is consistent with recent estimates of the 
mean free path to ionizing radiation (POW09).
}
\label{fig:literature}
\end{figure}

In Figure~\ref{fig:literature} we present estimates of $\ell(z)$
from several previous studies, parameterized as in Equation~\ref{eqn:powlaw}
(Table~\ref{tab:param} lists the parameters).  
The dotted lines show estimates from \cite{ssb} and \cite{lzt91},
which included very few observations at $z>3.5$ and 
had very discrepant guesses for the high $z$ universe.
The dashed line shows the result from \cite{key_lls} 
who integrated previous work 
with a new measurement at $z<1$ and an (still) unpublished survey
by Steidel \& Sargent.  Finally,
the solid curves show the results from \cite{storrie94} and \cite{peroux_dla03}
who surveyed LLSs at $z \sim 4$ using color-selected
quasars.  All of these analyses were claimed to correspond to 
the incidence of LLSs with $\mtll \ge 1$, \lzo,
although a careful review of the literature
raises doubts regarding this assertion.  Nevertheless, to make
comparisons with 
their reported $\mtll \ge 1$ results we have boosted each of
our \lzt\ estimations.
Formally, we estimate a correction of 6\%\ from our derived \fnhi\
distribution but, in practice, we adopt a more conservative 10\%\ correction.

\begin{deluxetable}{ccl}
\tablewidth{0pc}
\tablecaption{POWER-LAW PARAMETERS OF \lzo\ \label{tab:param}}
\tabletypesize{\footnotesize}
\tablehead{\colhead{\glls} & \colhead{$C_{\rm LLS}$} & \colhead{Reference}} 
\startdata
0.68 & 0.76 & \cite{ssb} \\
5.7  & 0.00081 & \cite{lzt91} \\
2.8  & 0.043   & \cite{storrie94} \\
1.5  & 0.25    & \cite{key_lls} \\
2.45 & 0.07    & \cite{peroux_dla03} \\
\enddata
\end{deluxetable}

Our results on \lzo\ at $z>4$ are lower than the values derived 
from the APM surveys, but within
$\approx 1\sigma$ of concordance.  The more important differences 
are between
the estimations at $z<4$.  All of the previous work was essentially
derived from the surveys of \cite{ssb} and \cite{lzt91} 
and, therefore, the curves all intersect at $z \approx 3$ at a 
\lzo~$\approx 2$.
Our results suggest that much of the previous work at $z \sim 3$ overestimated
the incidence of LLSs.  The original survey by \cite{ssb}
is in fair agreement with an extrapolation of our power-law form
for \lzo\ to $z=3$, but the reanalysis by \cite{key_lls} of 
unpublished spectra taken by Steidel \& Sargent led to a higher estimate
at this redshift.  We suspect that this later work gave \lzo\ values
that were too high either because of sample variance (i.e.\ small
number statistics), selection bias in the quasar sample, and/or
the systematic effects described in $\S$~\ref{sec:mock}.

It is also reasonable to consider whether our SDSS survey has
been biased low by an unidentified systematic error.
The measurement of \lmfp, however,  indicates that this is not the case.
Reconsider the analysis presented in $\S$~\ref{sec:mfp} (Figure~\ref{fig:kappa}).
If we adopted a 50\%\ higher incidence of $\mtll \ge 2$ LLSs, e.g.\
$\mltlls = 0.75$ at $z=3.7$, then we would infer a much steeper slope
for the LLSs ($\mbtlls \approx -1.5$) and then would require a
power-law shallower than $\beta = -0.5$ for 
$\mnhi = 10^{15} \cm{-2} - 10^{17} \cm{-2}$.
This would force \fnhi\ to steepen to $\beta < -3$ at 
$\mnhi \approx 10^{14.5} \cm{-2}$.  Such an extreme \fnhi\ distribution is
non-physical and, more importantly, ruled out by observed
line-statistics of the \lya\ forest \citep{kim02,misawa+07}.
We conclude that the incidence of LLSs at $z \approx 3.7$ cannot be
more than 30\%\ higher than our central value, and, at present,
we cannot identify a bias that would lead to such a large systematic 
underestimate in our results.

\subsection{Evolution in \lzo\ from $z=0-4$}
\label{sec:zevol}

Previous work has debated whether \lzo\ evolves
as a singe power-law $(1+z)^\mglls$ from $z \approx 0 - 4$
\citep[e.g.][]{key_lls}.
In $\S$~\ref{sec:loz}, we modeled our observations with a 
single power-law having $\mglls = \alls$.
The observed evolution in the mean free path (PWO09) also suggests
a steep evolution ($\mglls > 2$) for the LLSs.
We now consider whether a single power-law extrapolation is
a good description of \lzo\ for $z < 3$.

\begin{figure}[ht]
\centering
\includegraphics[height=3.5in,angle=90]{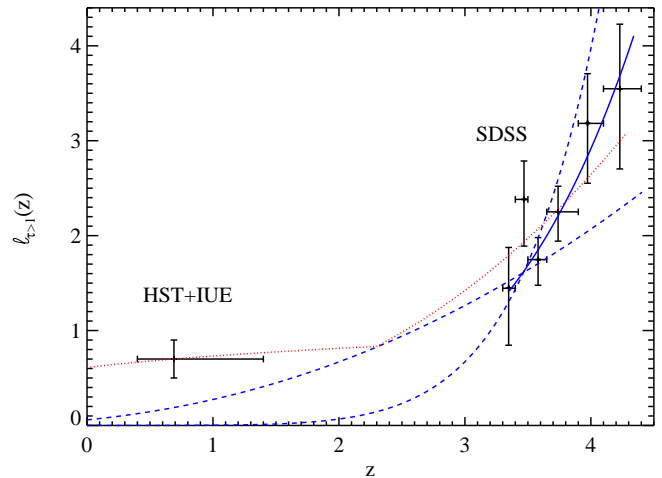}
\caption{A comparison of our results from the SDSS-DR7 survey of LLSs
(scaled to $\mtll \ge 1$) against the $z<1$ results from \cite{key_lls}.
The (blue) solid and dashed curves show the best-fit to the SDSS
observations and $\pm 2\sigma$ deviations from that fit.  None of
the curves, when extrapolated to $z=0$, intersect the low $z$ observations.
We rule out at high confidence that a single power-law $(1+z)^\mglls$
is a good description of the observations from $z=0-4$.
Instead, we suggest a break at $z \approx 2$, here modeled (dotted, red line)
as two power-laws with $\mglls = 2.78$ for $z>2.3$ and $\mglls = 0.26$
otherwise.
}
\label{fig:zloz}
\end{figure}

Survey of other \ion{H}{1} absorption systems have demonstrated
that a single $(1+z)^\gamma$ power-law is a poor description
of the \lya\ forest \citep{wjl+98} and the damped \lya\ systems
\citep{phh08}.  In the former case, one observes a flattening
in the \lya\ line-density at $z \sim 1$ which has been interpreted
to result from a corresponding decline in the intensity of 
the extragalactic UV background \citep[EUVB][]{wjl+98,dhk+99}.  
It is plausible that a similar
effect would influence the LLS.
In Figure~\ref{fig:zloz}, we present estimates of \lzo\
at $z \sim 3.5$, estimated by increasing the measured \lzt\ values by 10\%.
These are compared against the $z \sim 1$ measurement from 
\cite[][see Ribauldo et al., in prep for a new estimate]{key_lls}.
Overplotted on the data is a solid curve that shows the best-fit
power-law to the SDSS results.  The dashed curves show $2\sigma$
departures from this model, extrapolated to $z=0$.  
It is evident that none of these
curves intersect the low redshift observations.
We conclude, with high confidence, that a strict $(1+z)^\mglls$
power-law does not describe the evolution of \lzo\ from
$z=0-4$.   
Instead, the data suggest a `break' in the high $z$ power-law
at $z \sim 2$, similar to that observed for the \lya\ forest
although at somewhat higher redshift.  

\begin{figure*}
\centering
\includegraphics[height=6.5in,angle=90]{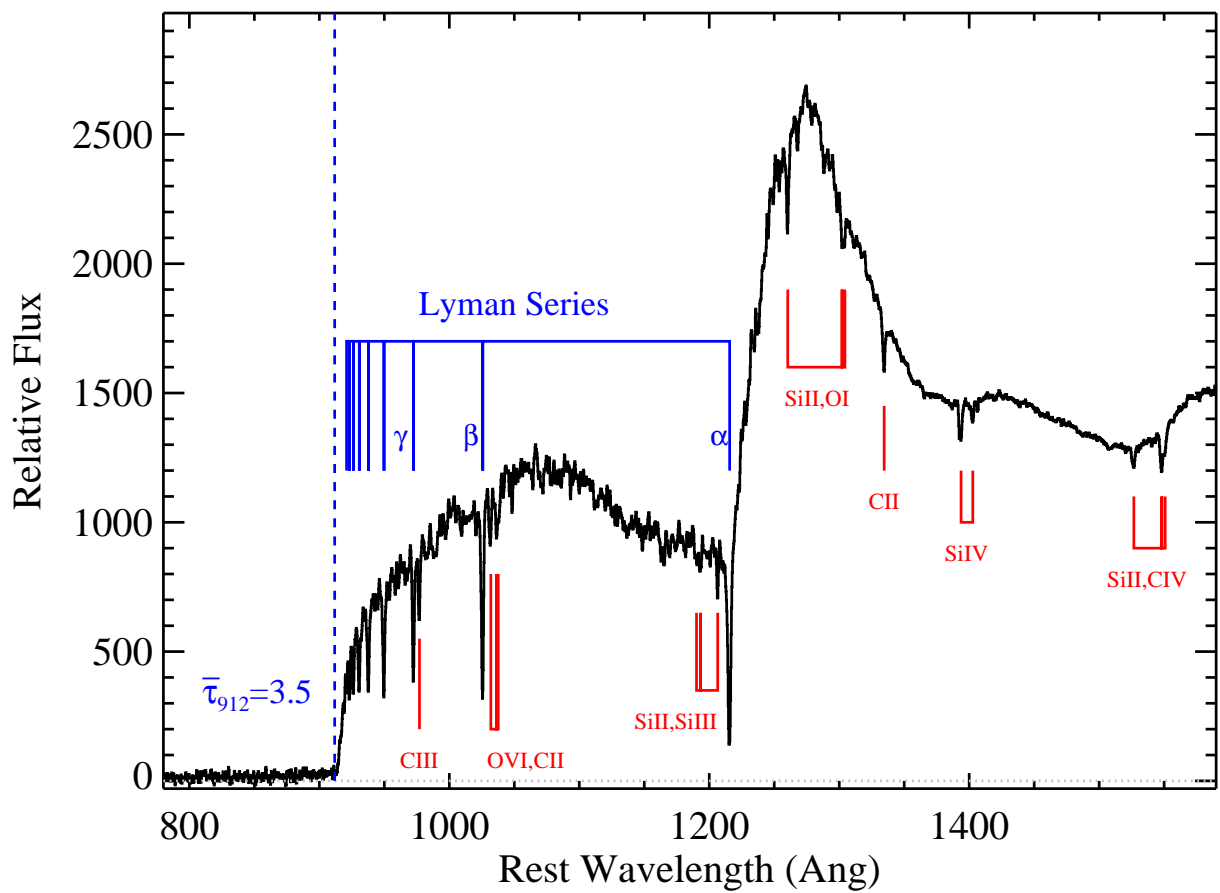}
\caption{The average absorption spectrum of $\mtll \ge 2$ LLSs
at $z \approx 3.7$.  Absorption from the \ion{H}{1} Lyman series is
readily apparent and one also notes strong Lyman limit absorption, as 
marked by the dashed vertical line.  The measured
average optical depth at the Lyman limit is $\mavtll \approx 3.5$, a
value that is inconsistent (too low) with our derived \fnhi\ distribution.
We infer that the stack includes the contribution from a small, but
non-negligible number of LLSs having $\mtll < 2$.  The
figure also identifies a series of metal-line transitions arising
from low and high-ions.  The presence of strong \ion{O}{6}
absorption is especially notable and suggests that LLSs
are comprised of gas in multiple phases.
}
\label{fig:stack}
\end{figure*}

The dotted curve, is an attempt to model this break.  
For $z<2.3$, we adopt the power-law form that matches
the \lya\ forest at low redshift ($\mglls = 0.26$) and
demand that it intersect the central value of the \cite{key_lls}
measurement.  For $z\ge 2.3$, the model breaks to a $\mglls = 2.78$
power-law, again consistent with the high $z$ evolution of the
\lya\ forest \citep[e.g.][]{kim02}.  This is a reasonably
good description of the data and we conclude that a break
in the power-law description of \lzo\ likely occurs at $z \approx 2$.
We will test this prediction with an (ongoing) survey for LLSs
at $z \sim 2$ in HST/ACS and WFC3 slitless spectra of 
$z \approx 2.3$ quasars (PI: O'Meara).

\subsection{The Average LLS Spectrum}
\label{sec:stack}

To gain additional insight into the absorption properties of the
LLSs, we have constructed an average (stacked) spectrum 
by (i) shifting each quasar spectrum containing a $\mtll \ge 2$ LLSs
to its rest-frame (\numlls\ systems total)
and (ii) averaging the fluxed data.
This stacked spectrum is primarily illustrative; 
it is shown in Figure~\ref{fig:stack}.  The peak in emission at 
$\lambda \approx 1280$\AA\ is from the \lya\ emission
peak of the background quasars.  The peak is 
offset from 1215\AA\ because the stack only includes intervening
LLSs, i.e.\ those that are offset by at least 3000 \kms\ from the 
quasar emission redshift.  
The peak's proximity to 1215\AA\ and relatively narrow width,
however, reflect that most LLSs in our survey are located within
$\delta z = 0.2$ of the quasar emission redshift.

The second strongest feature in the
spectrum is the Lyman limit at the expected wavelength of 912\AA.
Shortward of the Lyman limit, one observes a non-zero flux that
extends down to $\approx 770$\AA.  
We estimate the average optical depth of the Lyman limit absorption
in this stacked spectrum by assuming the flux at 980\AA\ provides a rough
estimate of the absorbed continuum at the Lyman limit and measure

\begin{equation}
\mavtll = -\ln \ltk \frac{f({\rm 900\AA})}{f({\rm 980\AA})} \rtk = 3.5 \perd
\end{equation}
We estimate a 20\%\ error in this value due to effects related to 
sky subtraction, uncertainty in the absorbed continuum $f(980\rAA)$, 
and the flux-weighted average of our stack.

The value of \avtll\ may be compared
to the average optical depth derived from our \fnhi\ distribution:

\begin{equation}
\mavtll = - \ln \ltb
\frac{\int \exp[-\mnhi \sigma_{LL}] \mfnhi \, d\mnhi}{\int \mfnhi \, d\mnhi}
\rtb
\label{eqn:avgtau}
\end{equation}
where the integrals are evaluated over the interval
$\mnhi = [10^{17.5},10^{22}] \cm{-2}$.
Evaluating at $z=3.7$ using the \fnhi\ distribution function shown
in Figure~\ref{fig:fn}, we derive
$\mavtll \approx 5.4$.
The frequency distribution in the LLS regime is sufficiently flat that
the higher \nhi\ systems (SLLS, DLAs) contribute significantly to the average.

The offset between these two evaluations is most likely due to the
inclusion of a non-negligible
number of systems having $\mtll < 2$.  Our analysis of mock spectra
and our internal comparison of the \tll\ estimates for the LLS
indicate that this occurs frequently.  
As described in $\S$~\ref{sec:lls},
we estimate a 0.2\,dex uncertainty in the \nhi\ values of LLSs
with $\mtll \approx 2$.
Although our tests also suggest this does not significantly
affect the estimate of \lzt, it 
can have a significant effect on the \avtll\ value
in the stacked spectrum.
We have repeated our calculation of \avtll\ extending the lower
limit of the \nhi\ distribution to $\mnhi = 10^{17.2} \cm{-2}$
instead of $\mnhi = 10^{17.5} \cm{-2}$.  For our favored \fnhi\
distribution, we calculate $\mavtll = 4.1$ and when allowing
for sample variance with a bootstrap analysis we find consisteny
in a non-negliglbe fraction of the trials ($>5\%$).

Returning to the stacked spectrum (Figure~\ref{fig:stack}), 
we note a series of absorption
lines corresponding to strong metal-line transitions
of low and high ionization ions.  This suggests a highly ionized, and
possibly multi-phase gas.  The  strong absorption of
\ion{O}{6}, in particular, suggests a multi-phase medium
consisting of at least one ``cool'' ($T \approx 10^4$K), 
photoionized phase and another,
more highly ionized phase which is presumably 
``warmer'' ($T \gtrsim 10^5$K) and possibly collisionally ionized.
This highly ionized phase has been detected in the damped \lya\
systems \citep{wp00a,fpl+07} and SLLSs \citep{fpl+07b} and its presence
in our average spectrum suggests it likely exists in lower
\nhi\ systems too \citep[e.g.][]{simcoe04}.
A more quantitative analysis of the metal-line absorption in LLSs,
however, awaits high-resolution spectroscopy \citep[e.g.][]{pks2000}.

\subsection{The Physical Nature of the LLSs}
\label{sec:nature}

As described in the introduction, observations and numerical
simulations associate the DLAs with high $z$ galaxies
residing in virialized dark matter halos \citep[e.g.][]{mwf+02,pgp+08}.  
The majority of absorption
lines comprising the \lya\ forest, meanwhile, are believed
to trace Mpc-scale overdensities in the medium between such
galaxies \citep[e.g.][]{mco+96}.  
By inference, one may associate LLSs with lower \nhi\
($< 10^{20} \cm{-2}$) with the interface between the IGM
and galaxies.  This inference, however, has not yet been 
extensively tested
by cosmological simulations or empirical observation.
Early works on the topic generally yielded too few LLSs in 
cosmological volumes \citep{kwh+96,gkh+01,mps+03}.
More recently, \cite{kg07} examined LLSs in a suite of simulations 
tuned to match the observed incidence of LLSs at $z \sim 4$.
Their simulations suggest LLSs are highly ionized gas
occupying volumes of space with dimension $1-100$\,kpc
and physically associated with galaxies of a wide range in mass.
The simulations, however, were not rigorously tested against
observations nor did they have sufficient spatial resolution
to ``establish the physical nature of these systems''.
The question remains: what is the physical nature of the LLSs?

Our observations place new constraints on the structures that
give rise to LLSs absorption.
The most informative measurements are \ltlls\ and the shape
of \fnhi\ in the LLSs regime.
Consider first \ltlls, which is proportional
to the comoving number density \nlls\ of the structures
times their average physical size \slls.  
In Figure~\ref{fig:complox}, we present the $\ell(X)$ values (in cumulative
form) for DLAs \citep{pw09} and SLLSs
\citep{opb+07}.
For the latter, we assume $\ell_{\rm SLLS}(X) = 0.20$ at all redshifts.
We have adopted a 20$\%$ lower (1$\sigma$) value than reported
by \cite{opb+07} to crudely correct for the SDSS quasar targeting bias
that will affect their measurement (PWO09). 
The figure demonstrates that the DLAs (especially) and 
the SLLSs have modest contributions to \ltlls.
At $z=3.4$, they contribute roughly half of the observed incidence
of $\mtll \ge 2$ LLS decreasing to
$\approx 33\%$ by $z=4$.  
This latter conclusion hinges on our assumptions for 
\lslls, in particular at $z \approx 4$ where the value is not
well constrained,  
but we expect the SLLSs to behave similarly
to the DLAs whose incidence is not increasing significantly at these
redshifts.  
We conclude that the incidence of $\mtll \lesssim 10$ LLSs \lzlls\ is
comparable to that of the DLAs and SLLSs.  As a result, it is
reasonable to associate all LLSs with the same structures,
i.e.\ gas within virialized halos as suggested by \cite{kg07}.
The principle challenge to this association is whether diffuse
halo gas has sufficiently high cross-section to LLS absorption.
In particular, one should consider whether the ``cold-flows''
identified in numerical simulations of galaxy formation
\citep{kkw+05,dbe+08} have sufficient density and size to
explain the majority of LLSs.  We are currently pursuing such analysis.

Now consider the evolution in \ltlls\ with redshift.
As noted in $\S$~\ref{sec:lox}, \ltlls\ is observed to decrease
with decreasing redshift.  Examining Figure~\ref{fig:complox} it
is evident that this decrease is driven by LLS with lower \tll,
i.e.\ by a significant decrease in \llls.   One possible explanation
for this decrease would be an increase in the intensity of the
EUVB with decreasing redshift, most likely due to higher emissivity from
the quasar population.  Unlike the DLAs and SLLSs systems
which have $\mtll > 100$ and therefore have regions that are self-shielded
from the EUVB, the $\mtll \lesssim 10$ LLSs are probably highly ionized
throughout \citep[e.g.][]{pks2000}.  Similar to the \lya\ forest,
the $\mtll \lesssim 10$ LLSs are sensitive to changes in the EUVB.
This interpretation would predict that the average
$U$ parameter for $\mtll \lesssim 10$ LLSs increase with decreasing redshift.
Another possible interpretation arises from linking LLS absorption
to the cold-flows in galactic halos predicted by numerical simulations.
The simulations reveal that the incidence of cold-flows in massive halos
$M \gtrsim 10^{12} \msun$ decreases with redshift \cite{dbe+08}, which
could result in a lower \ltlls\ for the LLSs.  These assertions warrant
further study and are likely to impact our understanding of the evolution
in the mean free path of the universe (PWO09).

Finally, consider the constraints on \lfnhi\ set by the observations.
As described in Figure~\ref{fig:fn}, we find that
\fnhi\ flattens at $\mnhi < 10^{19} \cm{-2}$ to a power-law
shallower than $\beta \approx -1$.  
The direct implication is that the cross-section of gas
 with $\mnhi = 10^{19} \cm{-2}$ {\it exceeds} that of
gas with $\mnhi = 10^{17.5} \cm{-2}$.  This is a remarkable
result.  Collapsed structures are generally observed to have
a density gradient where the highest density regions occupy a smaller
projected cross-section than lower density regions.  Our observations
indicate the opposite is true for LLSs with $\mtll \lesssim 10$.
This suggests that LLSs arise in 
 structures with an extended, higher surface
density region surrounded by a thin `layer' of gas with lower \nhi.
Such a description brings to mind the
cold-flows of accreting gas found in cosmological simulations of 
high $z$ galaxies \citep{kkw+05,dbe+08}.  Again, we plan to explore whether
such gas can explain 
the observed normalization and shape of \fnhi\ at $z>3.5$.

\section{Summary}
\label{sec:summary}

In this paper, we have performed a survey for $\mtll \ge 2$ LLS absorption
in the quasar spectra of the Sloan Digital Sky Survey, Data Release 7.
We established a spectral sample for statistical analysis
($\S$~\ref{sec:quasars}), estimated an absorbed continuum for each
quasar ($\S$~\ref{sec:continuum}), 
searched for LLSs using automated algorithms ($\S$~\ref{sec:lls}),
defined the survey path with strict criteria ($\S$~\ref{sec:path}),
and explored the effects of systematic bias and uncertainty with mock
spectra ($\S$~\ref{sec:mock}).
The primary results of this work are as follows:

\begin{enumerate}
\item We measure the incidence of $\mtll \ge 2$ LLSs \lzt\ at 
$z=3.3-4.4$ and find it is well modeled as a single power-law,
$\mlzt = C_{\rm LLS} [(1+z)/(1+z_*)]^\mglls$,
with $z_* \equiv 3.7$, $C_{\rm LLS} = \clls$, and
$\mglls = \alls$ (68\%\ c.l.).

\item A survey of LLSs in the SDSS spectra $\mzem < 3.6$ quasars confirms
a previously identified bias (PWO09) in the SDSS quasar targeting criteria
that biases the sample toward sightlines with foreground LLS absorption.

\item The number of $\mtll \ge 2$ LLS per unit absorption length \ltlls\
is observed to decrease by $\approx 50\%$ from $z=4$ to 3.4.
This indicates a decrease in the number of systems per comoving
Mpc$^3$ and/or a decrease in the average physical cross-section per system.
We suggest it is the latter effect, possibly related to an increase in the EUVB 
with decreasing redshift or a rising radiation field local to LLSs. 

\item The measured \ltlls\ values place an integral constraint on the \ion{H}{1}
frequency distribution \fnhi\ at $z \approx 3.7$.  Adopting previous estimates of \fnhi\ for 
$\mnhi \ge 10^{19} \cm{-2}$ \citep{opb+07,pw09}, we constrain
\fnhi\ for $\mnhi = 10^{17.5} - 10^{19} \cm{-2}$ assuming a power-law form
$\mlfnhi = k_{\rm LLS} \mnhi^{\beta_{\rm LLS}}$ to have
$k_{\rm LLS} \approx 10^{-4.5}$ and $\mbtlls = -0.8 \pm 0.3$.
This indicates a further
shallowing of the slope as one decreases \nhi\ below $10^{19} \cm{-2}$. 

\item Adopting constraints from the mean free path (PWO09) and \lya\ forest,
we derived new constraints on \fnhi\ at $z \approx 3.7$
for $\mnhi \approx  10^{15} - 10^{18} \cm{-2}$.
We find that $\beta \equiv d\ln\mfnhi/d\ln\mnhi$ must be steeper than $\beta = -1.5$
at $\mnhi \approx 10^{15} \cm{-2}$. 

\item We surveyed the spectra for proximate LLSs (PLLSs), those with redshifts
that are within 3000\kms\ of the quasar.  We measure an $\approx 25\%$ 
lower incidence of PLLSs than intervening systems at $z>3.5$.
This lends further support to the assertion that quasars have anisotropic
emission \citep{QPQ2}.
\end{enumerate}

Compared to previous work, our estimates of $\ell(z)$ show systematically
lower values.  We suggest that the difference is due to sample variance
and/or unidentified systematic bias in the prior analysis.
This conclusion is supported by measurements of the mean free path (PWO09)
which do not allow for a significantly higher incidence of LLSs.
We also find that the range of power-laws that describe our results at
$z \approx 3.5-4$ do not extrapolate to the results from $z<1$ observations
\citep{key_lls}.  We infer that the incidence of LLSs exhibits a break
at $z \approx 2$, qualitatively similar to that observed for the \lya\ forest
\citep[e.g.][]{wjl+98}.  The declining incidence of LLSs per
absorption length and the very shallow slope of \fnhi\ at $\mnhi < 10^{19} \cm{-2}$
suggest that $\mtll \lesssim 10$ LLSs arise in flattened (e.g.\ filamentary)
structures that have relatively sharp edges.  We associate these structures to 
the virialized halos that presumably give rise to SLLS and DLA absorption.
Finally, we encourage future work on whether such structures are consistent 
with the ``cold-flows'' identified
in numerical simulations. 

Through detailed analysis of biases and careful sample selection from
a large and homogeneous dataset,
this paper provides the first robust estimate of the incidence of 
LLSs at high redshift.
We note that the systematic errors described in $\S$~\ref{sec:mock} likely
limit the precision of any future $\ell(z)$ estimates to the order of $20-30\%$.
Nevertheless, this is sufficient to further explore the true evolution in $\ell(z)$
with redshift.  Programs with the Hubble Space Telescope for absorption
at $z<2$ and with ground-based observatories for $z>4$ are currently ongoing.
Altogether, these projects will describe the evolution of the UV background,
the growth of structure on galactic (and larger) scales,
and the chemical enrichment history of the universe.

\acknowledgments

The authors wish to recognize and acknowledge the very significant
cultural role and reverence that the summit of Mauna Kea has always
had within the indigenous Hawaiian community.  We are most fortunate
to have the opportunity to conduct observations from this mountain.
We thank P. Madau, M. Fumagalli, and C. Faucher-Giguerre 
for helpful comments.
This work was initiated in collaboration with S. Burles.
J. X. P. and J.M.O. are supported by NASA grants
HST-GO-10878.05-A and HST-GO-11594.01.  
J.X.P and G.W. are partially supported
by an NSF CAREER grant (AST--0548180) and by NSF grant AST-0908910.

Funding for the SDSS and SDSS-II has been provided by the Alfred P. Sloan 
Foundation, the Participating Institutions, the National Science Foundation, 
the U.S. Department of Energy, the National Aeronautics and Space 
Administration, the Japanese Monbukagakusho, the Max Planck Society, 
and the Higher Education Funding Council for England. The SDSS Web Site 
is http://www.sdss.org/.

The SDSS is managed by the Astrophysical Research Consortium for the Participating Institutions. The Participating Institutions are the American Museum of Natural History, Astrophysical Institute Potsdam, University of Basel, University of Cambridge, Case Western Reserve University, University of Chicago, Drexel University, Fermilab, the Institute for Advanced Study, the Japan Participation Group, Johns Hopkins University, the Joint Institute for Nuclear Astrophysics, the Kavli Institute for Particle Astrophysics and Cosmology, the Korean Scientist Group, the Chinese Academy of Sciences (LAMOST), Los Alamos National Laboratory, the Max-Planck-Institute for Astronomy (MPIA), the Max-Planck-Institute for Astrophysics (MPA), New Mexico State University, Ohio State University, University of Pittsburgh, University of Portsmouth, Princeton University, the United States Naval Observatory, and the University of Washington.

\appendix

\section{Comparisons with Keck+LRIS Spectra}
\label{appx:keck}

To assess uncertainties (statistical and systematic)
of surveying LLSs in the SDSS quasar spectra, we
obtained independent, higher quality spectra using the LRIS spectrometer
\citep{lris} on the Keck I telescope.  
LRIS employs a dichroic to split the data into two
spectral channels, each with its own camera.
For our observations, we employed the d560 dichroic which splits the 
light at $\approx 5600$\AA.
For the blue channel, we used the 640/4000 grism
which provides a dispersion of 0.63 \AA\ per unbinned pixel and has a nominal
wavelength coverage of $3100 \rAA < \lambda < 5600$ \AA.  The blue channel
data was binned by 2 in both the spatial and spectral dimensions. For the red
channel, we used the 600/7500 grating which provides a dispersion of
1.28 \AA\ per unbinned pixel, and which was tilted to provide a wavelength
coverage of $5600 \rAA < \lambda < 8200$\AA.  The red channel data was unbinned.
All observations were obtained using a 1 arc-second slit which provides
an $\approx 4$\,pixel FWHM corresponding to 
$\approx 290\mkms$ and $\approx 220 \mkms$ for the blue and red
data respectively.
The data were obtained in good sky conditions during a 4 night
run in October~2008 and had
exposure times ranging from 300 to 500 seconds.
The data were reduced using the LowRedux
pipeline\footnote{http://www.ucolick.org/$\sim$xavier/LowRedux/} which
bias subtracts, flat fields, optimally extracts, wavelength and flux
calibrates the data to produce a final 1D spectrum.  

\begin{figure*}
\centering
\includegraphics[height=6.5in,angle=90]{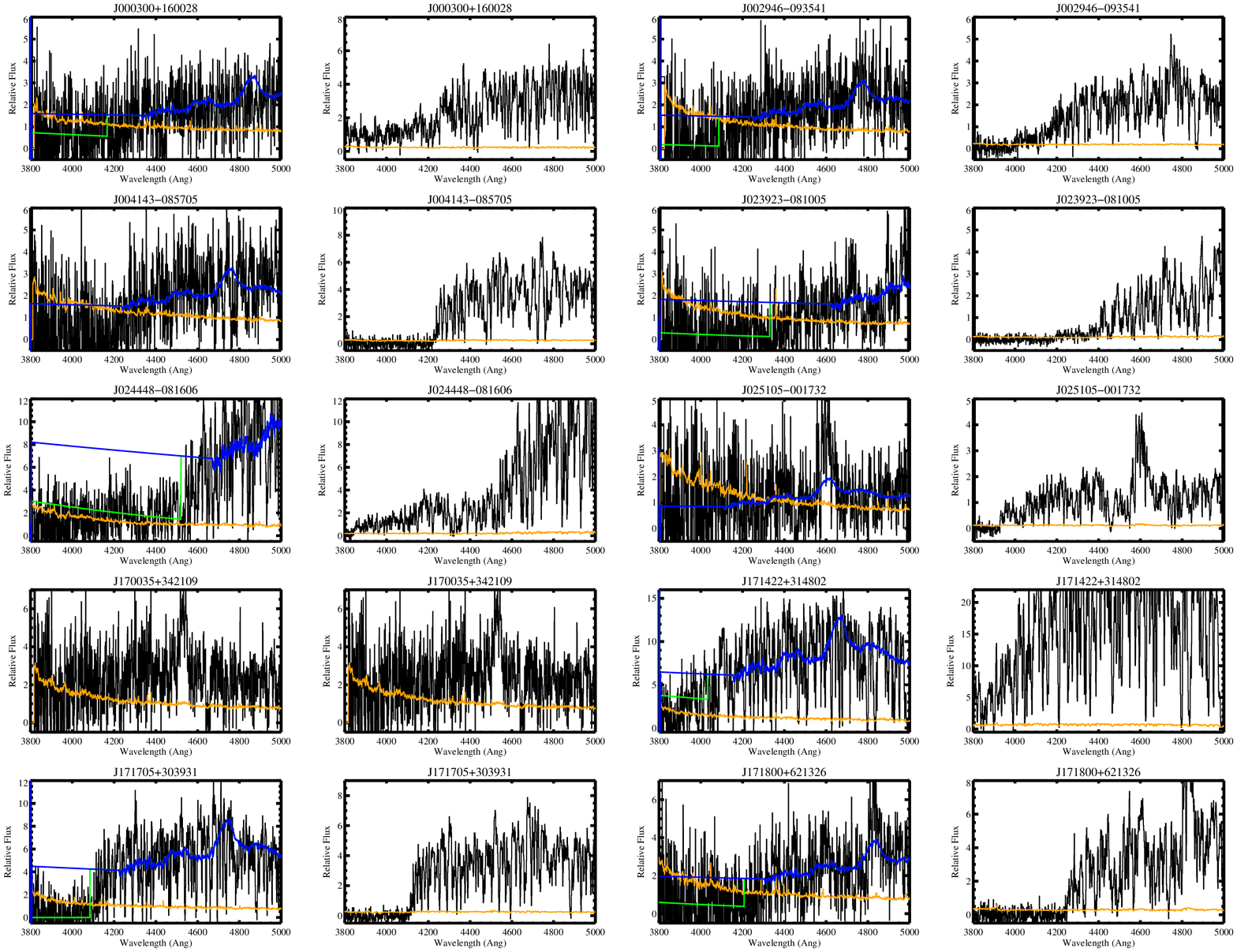}
\caption{Comparisons of SDSS (left panel in each pair) and
follow-up Keck/LRIS observations (right panel in each pair).
Overplotted on the SDSS spectrum is our model of the absorbed
continuum (blue) and the modeled foreground Lyman limit absorption (green).
For the SDSS spectra with \sna~$\ge 2$, we find excellent agreement
for analyses of each pair of spectra.
}
\label{fig:lris}
\end{figure*}

The SDSS targets for LRIS observations were chosen to sample a range
of LLSs, e.g., LLSs with $\mtll \ge 2$, pLLSs
candidates, PLLSs and spectra without apparent LLSs.  Furthermore, 
an emphasis was placed on quasars with lower S/N SDSS spectra 
to assess the completeness of recovering LLSs. 
In all cases, the LRIS spectra have sufficient S/N 
to unambiguously detect the presence of absorbers with $\mtll > 1$ 
over the full SDSS wavelength range (i.e.\ $\lambda > 3800$ \AA) for
the intervening LLS survey. 
Particular emphasis was given 
to determine what absorbed continuum S/N cutoff should be applied to
the SDSS sample.  To this end, two of the authors (JXP and JMO)
independently modeled LLS absorption in the LRIS data and 
compared the results to similar analysis of the 
the SDSS spectra.  The same
codes were used to model LLS absorption.  In Table
\ref{tab:lriscomp}, we present the results of these comparisons.  In
Figure \ref{fig:lris} we show a representative sample of the
LRIS data alongside their SDSS counterparts.  For the SDSS data in
Figure \ref{fig:lris}, we also show the continuum level assigned
to each spectrum.  

The LRIS/SDSS comparison illustrates that with a choice of absorbed
continuum S/N $>1$ 
we recover nearly 100\%\ of the LLSs with  $\mtll > 2$ in the SDSS.  We see this
explicitly in Table \ref{tab:lriscomp}, where we give the values for
\zstrt\ in the SDSS search, where \zstrt\ is the redshift at which
the absorbed continuum S/N crosses the value of 1.  In some cases, we
identify LLSs at redshifts lower than \zstrt\ in both the LRIS and
SDSS data.  Although these LLSs will not contribute to our results,
they lend additional confidence in our absorbed continuum S/N cutoff.
The only exceptions to the identification of LLSs in the LRIS and 
SDSS spectra at S/N~$\ge 1$ are for systems with $\mtll \approx 2$.
At this optical depth, JXP and JMO did not always
agree, even in the LRIS results. 
This highlights the fact that our results have an inherent uncertainty
in $\log \mnhi$ of $\approx 0.2$\,dex. 
Most importantly, this
disagreement does not appear to depend on the S/N of the spectrum for
the range of S/N we could expect from the SDSS data, and thus does not
effect our choice for the S/N threshold.



\begin{deluxetable}{lcccccccccc}
\tablewidth{0pc}
\tablecaption{SDSS-DR7 QUASARS WITH KECK+LRIS COMPARISON SPECTRA\label{tab:lriscomp}}
\tabletypesize{\footnotesize}
\tablehead{\colhead{Object Name} &\colhead{SDSS} &\colhead{SDSS} &\colhead{SDSS} &\colhead{SDSS} &\colhead{SDSS} &\colhead{SDSS} &\colhead{LRIS} &\colhead{LRIS} &\colhead{LRIS} &\colhead{LRIS}\\\colhead{} &\colhead{$z_{start,jxp}$} &\colhead{$z_{lls,jxp}$} &\colhead{log \nhi$_{jxp}$} &\colhead{$z_{start,jmo}$} &\colhead{$z_{lls,jmo}$} &\colhead{log \nhi$_{jmo}$} &\colhead{$z_{lls,jxp}$} &\colhead{log \nhi$_{jxp}$} &\colhead{$z_{lls,jmo}$} &\colhead{log \nhi$_{jmo}$}}
\startdata
J000300+160028&3.594&3.570&17.2&3.643&3.621&17.2&3.493&17.4&3.611&17.0\\
J002946-093541&3.411&3.523&$> 17.5$&3.682&3.482&$> 17.5$&3.563&17.4&3.524&17.4\\
J004143-085705&3.302&3.605&$> 17.5$&3.650&3.616&17.4&3.606&$> 17.5$&3.605&$> 17.5$\\
J023923-081005&0.000&0.000& 0.0&0.000&0.000& 0.0&3.827&$> 17.5$&3.801&17.4\\
J024448-081606&0.000&0.000& 0.0&3.280&3.957&17.2&3.949&17.4&3.980&17.4\\
J025105-001732&3.725&3.438&16.8&3.728&3.276&17.2&3.434&17.2&3.262&$> 17.5$\\
J170035+342109&0.000&0.000& 0.0&0.000&0.000& 0.0&3.064&$> 17.5$&3.061&$> 17.5$\\
J171422+314802&3.280&3.468&17.0&3.280&3.395&17.0&3.275&17.4&3.274&17.4\\
J171705+303931&3.280&3.480&$> 17.5$&3.280&3.475&$> 17.5$&3.476&$> 17.5$&3.477&$> 17.5$\\
J171800+621326&3.636&3.614&17.4&3.280&3.615&$> 17.5$&3.620&$> 17.5$&3.615&$> 17.5$\\
J173039+585847&0.000&0.000& 0.0&0.000&0.000& 0.0&2.776&$> 17.5$&2.775&$> 17.5$\\
J173115+563641&3.701&3.562&17.2&3.701&3.562&17.4&3.397&$> 17.5$&3.399&$> 17.5$\\
J204230-060112&0.000&0.000& 0.0&0.000&0.000& 0.0&3.862&17.4&3.863&$> 17.5$\\
J205142-071906&3.816&3.780&17.2&0.000&0.000& 0.0&3.799&$> 17.5$&3.796&17.4\\
J205509-071749&0.000&0.000& 0.0&3.280&3.859&$> 17.5$&3.553&$> 17.5$&3.550&$> 17.5$\\
J205551-004814&0.000&0.000& 0.0&0.000&0.000& 0.0&3.176&$> 17.5$&3.174&$> 17.5$\\
J210055-004843&3.538&3.331&$> 17.5$&3.597&3.357&$> 17.5$&3.331&$> 17.5$&3.329&$> 17.5$\\
J212204-001012&3.625&3.267&$> 17.5$&3.625&3.407&$> 17.5$&3.405&17.0&3.406&17.2\\
J212358-005350&3.280&3.626&$> 17.5$&3.280&3.627&$> 17.5$&3.626&$> 17.5$&3.626&$> 17.5$\\
J212444-005533&3.280&3.442&$> 17.5$&3.280&3.448&$> 17.5$&3.443&$> 17.5$&3.440&$> 17.5$\\
J214050+103832&3.754&3.737&17.2&3.760&3.745&17.2&3.705&17.2&3.687&17.2\\
J214227+005652&3.634&3.598&17.4&3.642&3.613&17.2&3.601&17.4&3.598&17.4\\
J220213-085222&3.280&0.000& 0.0&3.280&0.000& 0.0&3.144&$> 17.5$&3.143&$> 17.5$\\
J221014+114452&3.592&0.000& 0.0&3.592&3.304&$> 17.5$&3.286&$> 17.5$&3.285&$> 17.5$\\
J221458+135345&3.509&3.480&17.2&3.510&3.486&16.6&3.167&17.4&3.449&17.2\\
J222420-085339&3.660&3.577&17.4&3.660&3.577&17.4&3.480&17.0&3.021&$> 17.5$\\
J222824+134155&3.280&0.000& 0.0&3.302&3.944&16.8&3.725&17.2&3.442&17.0\\
J223659-080912&0.000&0.000& 0.0&0.000&0.000& 0.0&3.405&17.4&3.397&17.2\\
J224243-091544&0.000&0.000& 0.0&4.182&4.162&17.2&4.108&17.4&4.113&17.2\\
J224740-091512&0.000&0.000& 0.0&0.000&0.000& 0.0&4.175&$> 17.5$&3.895&16.0\\
J225053-084600&3.513&3.716&17.0&3.735&3.720&17.0&3.299&$> 17.5$&3.300&$> 17.5$\\
J225109-083138&3.280&3.867&$> 17.5$&0.000&0.000& 0.0&3.887&$> 17.5$&3.832&$> 17.5$\\
J225152+125707&3.547&3.386&17.2&3.547&3.386&17.4&3.360&$> 17.5$&3.360&$> 17.5$\\
J230022+125354&3.533&3.509&17.2&3.531&3.507&17.2&3.545&17.4&3.542&17.2\\
J230301-093931&3.280&3.316&$> 17.5$&3.280&3.312&$> 17.5$&3.311&$> 17.5$&3.308&$> 17.5$\\
J231137-084410&3.346&3.689&$> 17.5$&3.589&3.685&$> 17.5$&3.717&$> 17.5$&3.716&17.4\\
J232533+143247&0.000&0.000& 0.0&3.665&3.647&17.4&3.551&$> 17.5$&3.554&17.4\\
J233535-085939&3.643&3.621&17.2&3.640&3.620&17.2&3.389&$> 17.5$&3.335&$> 17.5$\\
J233634+133043&0.000&0.000& 0.0&0.000&0.000& 0.0&3.110&$> 17.5$&3.215&17.4\\
J234349-104742&3.551&3.403&$> 17.5$&3.551&3.367&$> 17.5$&3.366&$> 17.5$&3.363&$> 17.5$\\
\enddata
\tablecomments{{} List of all objects which have LRIS comparison spectra}
\end{deluxetable}
 
\clearpage


\end{document}